\documentclass[final,journal,letterpaper,twoside,twocolumn]{IEEEtran}

\usepackage[utf8]{inputenc} 
\usepackage[T1]{fontenc} 
\usepackage[cmex10]{amsmath} 

\usepackage{amssymb} 
\usepackage{mathtools} 
\usepackage{amsfonts} 
\usepackage{accents} 

\usepackage{mleftright}\mleftright 

\usepackage{multirow}

\usepackage[dvips]{graphicx} 
\graphicspath{{./}}
\DeclareGraphicsExtensions{.eps}

\usepackage[dvipsnames,table]{xcolor} 

\usepackage{bbm} 
\usepackage{xfrac} 
\usepackage{cancel} 
\usepackage{wrapfig} 
\usepackage{tensor} 

\usepackage{theorem}
\usepackage{cite}

\usepackage{url} 

\usepackage{psfrag} 

\usepackage[inline]{enumitem} 

\usepackage{array} 

\usepackage{aliascnt} 

\usepackage[hidelinks]{hyperref} 

\usepackage{cleveref} 

\usepackage{orcidlink} 

\usepackage[ruled,algoruled,vlined]{algorithm2e}

\SetCommentSty{mycommfont}

\usepackage{xparse} 

\usepackage{ifthen} 

\usepackage{comment}

\crefname{section}{Section}{Sections}
\crefname{subsection}{Section}{Sections}
\crefname{equation}{Eq.}{Equations}
\crefname{enumi}{part}{parts}
\crefname{table}{Table}{Tables}
\crefname{figure}{Figure}{Figures}
\crefname{algocf}{Algorithm}{Algorithms}

\newtheorem{theorem}{Theorem}
\crefname{theorem}{Theorem}{Theorems}

\newaliascnt{lemma}{theorem}
\newtheorem{lemma}[lemma]{Lemma}
\aliascntresetthe{lemma}
\crefname{lemma}{Lemma}{Lemmas}

\newaliascnt{definition}{theorem}
\newtheorem{definition}[definition]{Definition}
\aliascntresetthe{definition}
\crefname{definition}{Definition}{Definitions}

\newaliascnt{corollary}{theorem}
\newtheorem{corollary}[corollary]{Corollary}
\aliascntresetthe{corollary}
\crefname{corollary}{Corollary}{Corollarys}

\newaliascnt{claim}{theorem}

\aliascntresetthe{claim}
\crefname{claim}{Claim}{Claims}

\newaliascnt{conjecture}{theorem}

\aliascntresetthe{conjecture}
\crefname{conjecture}{Conjecture}{Conjectures}

\newaliascnt{question}{theorem}

\aliascntresetthe{question}
\crefname{question}{Question}{Questions}

\newaliascnt{example}{theorem}
\newtheorem{example}[example]{Example}
\aliascntresetthe{example}
\crefname{example}{Example}{Examples}

\newaliascnt{oquestion}{theorem}

\aliascntresetthe{oquestion}
\crefname{oquestion}{Open Question}{Open Questions}

\theoremstyle{plain}
\theorembodyfont{\normalfont}

\newaliascnt{remark}{theorem}

\aliascntresetthe{remark}
\crefname{remark}{Remark}{Remark}

\newtheorem{cnstr}{Construction}

\newenvironment{construction}{\begin{cnstr}}{\hfill$\Box$\end{cnstr}}
\crefname{cnstr}{Construction}{Constructions}

\crefname{step}{Step}{Steps}

\crefname{regime}{Regime}{Regimes}

\newtheorem{myalgo}{Algorithm}

\crefname{myalgo}{Algorithm}{Algorithms}

\newcounter{enumrom}
\renewcommand{\theenumrom}{(\roman{enumrom})}

\makeatletter
\renewcommand{\@endtheorem}{\endtrivlist}
\makeatother

\makeatletter
\renewcommand{\thefigure}{{\@arabic\c@figure}}
\renewcommand{\fnum@figure}{{\bf Figure\,\thefigure}}
\makeatother

\renewcommand{\leq}{\leqslant}
\renewcommand{\geq}{\geqslant}

\newcommand{\bfc}{{\boldsymbol c}}

\newcommand{\bfe}{{\boldsymbol e}}

\newcommand{\bfo}{{\boldsymbol o}}
\newcommand{\bfp}{{\boldsymbol p}}

\newcommand{\bfu}{{\boldsymbol u}}
\newcommand{\bfv}{{\boldsymbol v}}
\newcommand{\bfw}{{\boldsymbol w}}
\newcommand{\bfx}{{\boldsymbol x}}
\newcommand{\bfy}{{\boldsymbol y}}
\newcommand{\bfz}{{\boldsymbol z}}

\newcommand{\bfzero}{{\boldsymbol 0}}

\newcommand{\cB}{\mathcal{B}}
\newcommand{\cC}{\mathcal{C}}
\newcommand{\cD}{\mathcal{D}}

\newcommand{\cO}{\mathcal{O}}
\newcommand{\cP}{\mathcal{P}}

\newcommand{\cT}{\mathcal{T}}
\newcommand{\cU}{\mathcal{U}}

\newcommand{\cX}{\mathcal{X}}

\newcommand{\cZ}{\mathcal{Z}}

\renewcommand{\Bbb}{\mathbb}

\newcommand{\N}{{\Bbb N}}

\newcommand{\1}{\mathbbm{1}}

\newcommand{\vcomment}[1]{{\color{violet}#1}}

\newcommand{\ecomment}[1]{{\color{Emerald}#1}}

\newcommand{\ncomment}[1]{{\color{NavyBlue}#1}}

\newcommand{\yy}[1]{{\footnotesize [\ncomment{#1}\;\;\vcomment{--Yoni}]}}

\newcommand{\db}[1]{{\footnotesize [\ecomment{#1}\;\;\vcomment{--Daniella}]}}

\DeclarePairedDelimiter\abs{\lvert}{\rvert}
\DeclarePairedDelimiter\ceilenv{\lceil}{\rceil}
\DeclarePairedDelimiter\floorenv{\lfloor}{\rfloor}
\DeclarePairedDelimiter\parenv{\lparen}{\rparen}
\DeclarePairedDelimiter\sparenv{\lbrack}{\rbrack}
\DeclarePairedDelimiter\bracenv{\lbrace}{\rbrace}
\DeclarePairedDelimiter\multbrace{\lbrace\!\!\lbrace}{\rbrace\!\!\rbrace}
\DeclarePairedDelimiterX\mathset[2]{\lbrace}{\rbrace}{#1 : #2}
\DeclarePairedDelimiterX\multset[2]{\lbrace\!\!\lbrace}{\rbrace\!\!\rbrace}{#1 : #2}
\DeclarePairedDelimiterX\inner[2]{\langle}{\rangle}{#1 \mathrel{},\mathrel{} #2}
\DeclarePairedDelimiterX\condparenv[2]{(}{)}{#1 \mathrel{}\delimsize\vert\mathrel{} #2}

\DeclareDocumentCommand\norm{ o m }{
    \IfNoValueTF{#1}
        {\left\Vert#2\right\Vert}
        {\left\Vert#2\right\Vert_{#1}}
}

\DeclareDocumentCommand\der{ o m o }{
    \IfNoValueTF{#1}
        {
            \IfNoValueTF{#3}
                {\frac{d}{d{#2}}}
                {\frac{d{#3}}{d{#2}}}
        }
        {\parenv*{\frac{d}{d{#2}}}^{#1}\IfNoValueTF{#3}{}{#3}}
}
\DeclareDocumentCommand\partder{ o m m }{
    \IfNoValueTF{#1}
        {\frac{\partial{#3}}{\partial{#2}}}
        {\frac{\partial^{#1}{#3}}{{\partial{#2}}^{#1}}}
}
\DeclareDocumentCommand\df{ o m o }{
    d\IfNoValueTF{#1}{}{^{#1}}{#2}\IfNoValueTF{#3}{}{_{#3}}
}

\newcommand{\deq}{\mathrel{\triangleq}}

\DeclareMathOperator{\supp}{supp}

\DeclareMathOperator{\ind}{Ind}

\DeclareMathOperator{\red}{red}

\newcommand{\lmin}{L_{\min}}
\newcommand{\lmax}{L_{\max}}
\newcommand{\whlmax}{\widehat{L}_{\max}}

\newcommand{\lmea}{L_{\rm mean}}

\newcommand{\per}[1]{\perp^{\!\! #1}}

\DeclareDocumentCommand\enc{ o }{
    \IfNoValueTF{#1}
        {\operatorname{Enc}}
        {\operatorname{Enc}_{\ref*{#1}}}
}

\DeclareDocumentCommand\dec{ o }{
    \IfNoValueTF{#1}
        {\operatorname{Dec}}
        {\operatorname{Dec}_{\ref*{#1}}}
}

\makeatletter
\newcommand\code[1]{%
  \@ifundefined{r@#1}{%
    \cC_{\operatorname*{#1}}%
  }{%
    \cC_{\ref*{#1}}%
  }%
}
\makeatother

\newcommand{\ball}[1]{\cB_{\operatorname*{#1}}}

\begin{document}

\title{
Adversarial Torn-paper Codes}

\author{Daniella~Bar-Lev\,\orcidlink{0000-0001-6766-1450}\,%
		,~\IEEEmembership{Student~Member,~IEEE}, 
		Sagi~Marcovich\,\orcidlink{0000-0003-4165-2024}\,%
		,~\IEEEmembership{Student~Member,~IEEE} 
		Eitan~Yaakobi\,\orcidlink{0000-0002-9851-5234}\,%
		,~\IEEEmembership{Senior~Member,~IEEE}, 
		and Yonatan~Yehezkeally\,\orcidlink{0000-0003-1652-9761}\,%
		,~\IEEEmembership{Member,~IEEE}
  \thanks{%
  Manuscript received 26~August~2022; revised 20~February~2023; accepted 29~June~2023. 
    This work was supported in part by the European Research Council (ERC) under the European Union's Horizon~2020 Research and Innovation Program under Grant~801434, and in part by the European Union (ERC, DNAStorage, 865630). Views and opinions expressed are however those of the authors only and do not necessarily reflect those of the European Union or the European Research Council Executive Agency. Neither the European Union nor the granting authority can be held responsible for them. %
  The work of Daniella~Bar-Lev, Sagi~Marcovich, and Eitan~Yaakobi was supported in part by the U.S.-Israel Binational Science Foundation (BSF) under Grant~2018048. 
  The work of Yonatan~Yehezkeally was supported the Alexander von Humboldt Foundation under a Carl Friedrich von Siemens Post-Doctoral Research Fellowship. 
  An earlier version of this paper was presented in part at the 2022 {IEEE} International Symposium on Information Theory ({ISIT}) [\textsc{DOI:\; 10.1109/ISIT50566.2022.9834766}]. 
  \emph{(Daniella Bar-Lev, Sagi Marcovich, and Yonatan Yehezkeally contributed equally to this work.)}
  \emph{(Corresponding author: Yonatan Yehezkeally.)}}
  \thanks{%
  Daniella~Bar-Lev, Sagi~Marcovich, and Eitan~Yaakobi are with the Department of Computer Science, Technion---Israel Institute of Technology, Haifa 3200003, Israel 
  (e-mail: \texttt{daniellalev@cs.technion.ac.il}; \texttt{sagimar@cs.technion.ac.il}; \texttt{yaakobi@cs.technion.ac.il}).
  Yonatan~Yehezkeally is with the Institute for Communications Engineering, School of Computation, Information and Technology, Technical University of Munich, 80333 Munich, Germany 
  (e-mail: \texttt{yonatan.yehezkeally@tum.de}).}
  \thanks{Copyright (c) 2023 IEEE. Personal use of this material is permitted. Permission from IEEE must be obtained for all other uses, in any current or future media, including reprinting/republishing this material for advertising or promotional purposes, creating new collective works, for resale or redistribution to servers or lists, or reuse of any copyrighted component of this work in other works.}
}

\maketitle
\begin{abstract}
We study the \emph{adversarial torn-paper channel}. This problem is 
motivated by applications in DNA data storage where the DNA strands 
that carry information may break into smaller pieces which are 
received out of order. Our model extends the previously researched 
probabilistic setting to the worst-case. We develop code constructions 
for any parameters of the channel for which non-vanishing asymptotic 
rate is possible and show our constructions achieve asymptotically 
optimal rate while allowing for efficient encoding and decoding. 
Finally, we extend our results to related settings included 
multi-strand storage, presence of substitution errors, or incomplete 
coverage.
\end{abstract}

\section{Introduction} \label{sec:intro}

High density and extreme longevity make DNA an appealing medium for 
data storage, especially for archival purposes~\cite{WonWonFoo03, 
ChuGaoKos12, Bal13, dnaalliancewp21}. Advances in DNA synthesis and 
sequencing technologies and recent proofs of concept~\cite{ChuGaoKos12, 
GolCheDesLePSipBir13, GraHecPudPauSta15, BorLopCarCezSeeStr16, 
ErlZie17, OrgAngCheetal18} have ignited active research into the 
capacity and challenges of data storage in this medium. 

An aspect of this medium is that typically only short DNA sequences 
may be read; information molecules are therefore broken up into 
pieces and then read out of order, such as in shotgun 
sequencing~\cite{MotBreTse13, BreBreTse13, GabMil19, RavVahSho22}. 
Multiple channel models have recently been suggested and studied 
based on this property. An assumption of overlap in read substrings 
and (near) uniform coverage leads to the problem of string 
reconstruction from substring composition \cite{AchDasMilOrlPan15, 
MotBreTse13, MotRamTseMa13, GanMosRac16, BreBreTse13, ShoCouTse15, 
ShoKamGovXiaCouTse16}; on the contrary, assuming no overlap in read 
substrings leads to the \emph{torn-paper problem}~\cite{RavVahSho21, 
ShoVah21, NasShoVah22}, a problem closely related to the shuffling 
channel~\cite{ShoHec19, HecShoRamTse17,LenSieWacYaa19,WeiMer22}. 
This problem is motivated by DNA-based storage systems, where the 
information is stored in synthesized strands of DNA molecules. 
However, during and after synthesis, the DNA strands may break into 
smaller segments and due to the lack of ordering among the 
strands in these systems, all broken segments can only be read 
out of order~\cite{ShoVah21}. Thus, the goal is to successfully 
retrieve the data from this collection of read segments of the broken 
DNA strands.

In the \emph{torn-paper channel}~\cite{RavVahSho21, ShoVah21}, also 
known as the \emph{chop-and-shuffle channel}~\cite{NasShoVah22}, a 
long information string is segmented into non-overlapping substrings 
and their length has some known distribution. The channel outputs an 
unordered collection of these substrings, preserving their 
left-to-right orientation. 
Given the lengths' distribution, the goal is to determine the channel 
capacity and devise efficient coding techniques. The geometric 
distribution was first studied in~\cite{ShoVah21}, and later 
in~\cite{NasShoVah22} using the Varshamov-Tenengolts (VT) 
codes~\cite{VarTen65}. Subsequently,  \cite{RavVahSho21} considered 
almost arbitrary distributions while, additionally, extending the 
problem by introducing incomplete coverage, i.e., assuming some of the 
substrings are deleted with some probability. 

The torn-paper channel was studied so far only in the probabilistic 
setting. The goal of this paper is to extend this channel to the worst 
case, referred to herein as the \emph{adversarial torn-paper channel}. 
Namely, it is assumed that an information string is adversarially 
segmented into non-overlapping substrings, where the length of each 
substring is between $\lmin$ and $\lmax$, for some given $\lmin$ and 
$\lmax$. 
We show that the capacity of this channel is determined by $\lmin$, 
whereas the capacity of the probabilistic channel was shown to depend 
on the \emph{average} substring length; nevertheless, we choose this 
adversarial model here for ease of analysis, and observe that under 
this setting the average substring length might indeed approach 
$\lmin$. For further discussion of an average-restricted adversary, 
see~\cref{sec:conc}.

We study the noiseless adversarial torn-paper channel for a single 
information string, as well as multiple strings, which is motivated by 
DNA sequencing technologies where multiple strings are sequenced 
simultaneously~\cite{Chinetal13,LomQuiSim15,Sal10}. We also extend the 
model to either allow for substitution errors affecting the 
information string prior to segmentation, or for incomplete coverage 
due to deletion of several segments after the segmentation. In all 
cases we investigate the values of $\lmin$ and $\lmax$ that permit 
codes with non-vanishing asymptotic rates, and develop constructions 
of codes with efficient encoding and decoding algorithms, 
asymptotically achieving optimal rates. 

The rest of this paper is organized as follows. In~\cref{sec:def}, the 
definitions and notations that will be used throughout the paper are 
presented, as well as a lower bound on $\lmin$ required for the 
existence of codes for the adversarial torn-paper channel with 
non-vanishing asymptotic rates. In~\cref{sec:torn-single} we first 
study the application of a known code construction to the adversarial 
channel, and observe its limitations in that setting; then, we present 
the basic construction used throughout the paper for the noiseless 
case of the single-strand adversarial torn-paper channel, and extend 
it to the multi-strand case. In~\cref{sec:torn-single-sub} we extend 
our construction to two noisy settings, including substitution errors 
or incomplete coverage. We conclude with a summary and remarks 
in~\cref{sec:conc}.

\section{Definitions and Preliminaries}\label{sec:def}

Let $\Sigma$ be a finite alphabet of size $q$. 
For convenience of presentation, we assume $\Sigma$ is equipped with a 
ring structure, and in particular identify elements $0,1\in \Sigma$. 
For a positive integer $n$, let $[n]$ denote the set $[n]\deq 
\bracenv*{0, 1, \ldots, n-1}$. Let $\Sigma^*$ denote the set of all 
finite strings over $\Sigma$. The length of a string~$\bfx\in 
\Sigma^*$ is denoted by $\abs*{\bfx}$. We also denote, for $\bfx = 
(x_i)_{i\in [n]} \in \Sigma^n$, its \emph{support} $\supp(\bfx)\deq 
\mathset*{i\in [n]}{x_i\neq 0}$, and $\norm{\bfx}\deq 
\abs*{\supp(\bfx)}$. For strings $\bfx,\bfy\in\Sigma^*$, we denote 
their concatenation by $\bfx\circ\bfy$. We say that $\bfv$ is a 
\emph{substring}, or \emph{segment}, of~$\bfx$ if there exist strings 
$\bfu,\bfw$ (perhaps empty) such that $\bfx = \bfu\circ \bfv\circ 
\bfw$. If $\abs*{\bfv} = \ell$, we specifically say that $\bfv$ is an 
\emph{$\ell$-substring} (\emph{$\ell$-segment}) of $\bfx$. If 
$\abs*{\bfu}=i$ then it is said that 
$\bfv$ is the substring (similarly, $\ell$-substring) of 
$\bfx$ at \emph{location}~$i$. 
We say that $\bfv$ appears \emph{cyclically} in~$\bfx$, at 
location~$i$, if $\bfx = \bfu\circ \bfw$ and $\bfv$ is the substring 
of $\bfw\circ \bfu$ at location~$(i-\abs*{\bfu})$. 
For example, $010$ is the $3$-substring of~$00101$ at location~$1$, and 
also its $3$-substring at location~$3$, where the latter is a cyclic 
appearance. 
We avoid using the term \emph{index} as it is reserved to elements of 
presented constructions.

In our setting, information is stored in an unordered collection of 
strings over $\Sigma$; it might be allowed for the same string to 
appear with multiplicity in the collection, which is encapsulated in 
the following formal definition: 
\begin{align*}
\cX_{n,k} &\deq \mathset*{S=\multbrace*{\bfx_0,\ldots,
\bfx_{k-1}}}{\forall i, \bfx_i\in \Sigma^n}.
\end{align*}
Here, $\multbrace*{a,a,b,\ldots}$ denotes a multiset; i.e., 
elements appear with multiplicity (but no order). Note that 
$\abs*{\cX_{n,k}} = \binom{k+q^n-1}{k}$. 
It is assumed that a message $S\in\cX_{n,k}$ is read by segmenting all 
elements of $S$ into non-overlapping substrings of lengths between 
some fixed values $\lmin$ and $\lmax$, and all segments are received, 
possibly with multiplicity, without order or information on which 
element they originated from. 
More formally, a \emph{segmentation} of the string $\bfx$ is a 
multiset $\multbrace*{\bfu_0,\bfu_1,\ldots,\bfu_{m-1}}$, 
where $\bfx$ can be presented as $\bfx = \bfu_0\circ \bfu_1\circ 
\cdots\circ \bfu_{m-1}$. In case $\lmin\leq \abs*{\bfu_i}\leq \lmax$ 
for $0\leq i <m-1$ and $\abs*{\bfu_{m-1}}\leq \lmax$, then the 
segmentation is called an \emph{$(\lmin,\lmax)$-segmentation}. The 
set of all $(\lmin,\lmax)$-segmentations of $\bfx$ is denoted by 
$\cT_{\lmin}^{\lmax}(\bfx)$ and is referred as the \emph{$(\lmin,
\lmax)$-segmentation spectrum of $\bfx$}. 
For example, 
\begin{align*}
	\cT_2^3(00101) &= 
	\bracenv[\Big]{\multbrace*{001,01}, 
	\multbrace*{00,101}, 
	\multbrace*{00,10,1}}.
\end{align*}
These definitions are 
naturally extended for a multiset $S\in \cX_{n,k}$, so a 
\emph{segmentation} of $S$ is a union (as a multiset) of segmentations 
of all the strings in $S$ (and the same holds for an $(\lmin, 
\lmax)$-segmentation), 
and $\cT_{\lmin}^{\lmax}(S)$, the 
\emph{$(\lmin,\lmax)$-segmentation spectrum of $S$}, is the set of all 
$(\lmin,\lmax)$-segmentations of $S$.

Note that our channel model only restricts the length of the last 
segment to be at most $\lmax$. Such a relaxation is motivated in 
applications where segmentation of the strings occurs sequentially, so 
that it might happen that the last segment is shorter than $\lmin$, 
but not larger than $\lmax$. 

A code $\cC\subseteq\cX_{n,k}$ is said to be an \emph{$(\lmin, 
\lmax)$-multistrand torn-paper code} if for all $S,S'\in\cC$, $S\neq 
S'$, it holds that all possible $(\lmin,\lmax)$-segmentations of 
$S,S'$ are distinct. That is, $\cT_{\lmin}^{\lmax}(S) \cap 
\cT_{\lmin}^{\lmax}(S') =\emptyset$. For $k=1$, we simply refer to 
\emph{$(\lmin,\lmax)$-single strand torn-paper codes}. 

In case $\lmin = \lmax = \ell$, then for convenience, we let 
$\cT_\ell(\bfx)\deq \cT_\ell^\ell\parenv*{\bfx}$ and $\cT_{\ell}(S) 
\deq \cT_\ell^\ell(S)$ and note that in this case 
$\abs*{\cT_{\ell}(\bfx)} = \abs*{\cT_{\ell}(S)}=1$. 
For example, if $S = \multbrace*{01010,00101,11101}$ (which 
may be thought of as a multiset), then
\begin{align*}
    \cT_2(S) = 
    \bracenv[\Big]{\multbrace*{01,01,0,00,10,1,11,10,1}}.
\end{align*}

Note that $\cT_\ell(S)$ is only one possible channel output given 
input $S$. Nevertheless, $\cT_{\lmin}(S) \subseteq 
\cT_{\lmin}^{\lmax}(S)$ for all $S$ and $\lmin\leq\lmax$, hence 
every $(\lmin,\lmax)$-multistrand torn-paper code $\cC\subseteq 
\cX_{n,k}$ satisfies 
\begin{align}\label{eq:trivialbound}
    \abs*{\cC}\leq \abs*{\mathset*{\cT_{\lmin}(S)}{S\in\cX_{n,k}}}.
\end{align}

For all $\cC\subseteq \cX_{n,k}$ we denote the \emph{rate}, 
\emph{redundancy} of~$\cC$ by $R(\cC)\deq 
\frac{\log\abs*{\cC}}{\log\abs*{\cX_{n,k}}}$, $\red(\cC)\deq 
\log\abs*{\cX_{n,k}}-\log\abs*{\cC}$, respectively. 
Throughout the paper, we use the base-$q$ logarithms.

For two non-negative functions~$f,g$ of a common variable~$n$, 
denoting $L\deq \limsup_{n\to\infty}\frac{f(n)}{g(n)}$ (in the wide 
sense, i.e., $L=\infty$ if $\frac{f(n)}{g(n)}$ is unbounded) we say 
that $f=o_n(g)$ if $L=0$, $f=\Omega_n(g)$ if $L>0$, $f=O_n(g)$ if 
$L<\infty$, and $f=\omega_n(g)$ if $L=\infty$. If $f$ is not positive, 
we say $f=O_n(g)$ ($f=o_n(g)$) if $\abs*{f} = O_n(g))$ (respectively, 
$\abs*{f}=o_n(g)$). We say that $f=\Theta_n(g)$ if $f=\Omega_n(g)$ and 
$f=O_n(g)$. If clear from context, we omit the subscript from 
aforementioned notations.

We conclude this section by observing a lower bound on the required 
segment length $\lmin$ for multi-strand torn-paper codes to achieve 
non-vanishing rates, and in particular rates approaching one.
\begin{lemma}\label{lem:torn-lin-red}
	If $\log(k)=o(n)$ and $\lmin = a \log(n k) + O_{n k}(1)$ for 
	some $a\geq 1$, then 
	\begin{align*}
		&\log\abs*{\cX_{n,k}} - 
		\log\abs*{\mathset*{\cT_{\lmin}(S)}{S\in\cX_{n,k}}} \\
		&\quad\geq n k \parenv*{\frac{1}{a} - a \frac{\log(k)}{n} 
		- O\parenv*{\frac{\log\log(n k)}{\log(n k)}}}.
	\end{align*}
\end{lemma}
\begin{IEEEproof}
First, note that 
\begin{align*}
	\abs*{\cX_{n,k}} = \binom{k+q^n-1}{k}\geq \frac{q^{nk}}{k!}
	\geq \frac{q^{nk}}{k^k},
\end{align*}
and hence $\log\abs*{\cX_{n,k}} \geq (n-\log(k)) k$. 
Next, since $\abs*{\mathset*{\cT_{\lmin}(S)}{S\in\cX_{n,k}}}$ is 
monotonically non-decreasing in~$n$, we have that 
\begin{align*}
	\abs*{\mathset*{\cT_{\lmin}(S)}{S\in\cX_{n,k}}} 
	&\leq \binom{k\ceilenv*{n/\lmin}+q^{\lmin}-1}{q^{\lmin}-1} \\
	&\leq \binom{k\ceilenv*{n/\lmin}+q^{\lmin}}{q^{\lmin}}.
\end{align*}

Now, for $v\geq u\geq 0$ we observe 
\begin{IEEEeqnarray*}{+rCl+x*}
	\log\binom{u+v}{u} 
	&\leq& \log\frac{1}{u!}(u+v)^u 
	\leq u \log\parenv*{e (1 + \frac{v}{u})} \\
	&\leq& u \parenv*{\parenv*{1+\frac{u}{v}}\log(e) + 
	\log(\frac{v}{u})} \\
	&\leq& u (2 \log(e) + \log(\frac{v}{u})), 
\end{IEEEeqnarray*}
where we used $\log(1+x)\leq \frac{\log(e)}{x}+\log(x)$. 
Setting $u\deq k\ceilenv*{n/\lmin} \leq \frac{n k}{\lmin} + k$ and 
$v\deq q^{\lmin} = \Theta\parenv*{(n k)^a}$, we have $\frac{v}{u} = 
\Theta\parenv*{(n k)^{a-1} \lmin} = \Theta\parenv*{(n k)^{a-1} \log(n 
k)}$, and therefore $\log\parenv*{\frac{v}{u}} = (a-1)\log(n k) + 
\log\log(n k) + O(1)$. 
We then conclude
\begin{IEEEeqnarray*}{+rCl+x*}
	\IEEEeqnarraymulticol{3}{l}{
	\log\abs*{\mathset*{\cT_{\lmin}(S)}{S\in\cX_{n,k}}} 
	} \\* 
	\quad &\leq& \parenv*{\frac{n k}{\lmin} + k} 
	\big((a-1)\log(n k) + \log\log(n k) \>+ \\*
	&& \IEEEeqnarraymulticol{1}{r}{O(1)\big)} \\
	&=& (a-1) \frac{n k \log(n k)}{\lmin} + k (a-1)\log(n k) \>+ \\*
	&& \IEEEeqnarraymulticol{1}{r}{\parenv*{\frac{n k}{\lmin} + k} 
	\parenv*{\log\log(n k) + O(1)}} \\
	&=& (a-1) \frac{n k \log(n k)}{\lmin} + k (a-1)\log(n k) \>+ \\*
	&& \IEEEeqnarraymulticol{1}{r}{O\parenv*{\frac{n k \log\log(n k)}{\log(n k)}}} \\
	&=& n k \bigg((a-1) \frac{\log(n k)}{\lmin} 
	+ (a-1) \frac{\log(k)}{n} \>+ \\*
	&& \IEEEeqnarraymulticol{1}{r}{O\parenv*{\frac{\log(n)}{n}} + 
	O\parenv*{\frac{\log\log(n k)}{\log(n k)}} \bigg)} \\
	&=& n k \bigg(\frac{a-1}{a + O\parenv*{1/\log(n k)}} 
	+ (a-1) \frac{\log(k)}{n} \>+ \\*
	&& \IEEEeqnarraymulticol{1}{r}{O\parenv*{\frac{\log\log(n k)}{\log(n k)}}\bigg)} \\
	&=& n k \parenv*{\frac{a-1}{a} + (a-1) \frac{\log(k)}{n} 
	+ O\parenv*{\frac{\log\log(n k)}{\log(n k)}}},
\end{IEEEeqnarray*}
which verifies the lemma's statement. 
\end{IEEEproof}
We note that throughout this paper, we perform redundancy analysis to 
the second-most-significant term, and retain the order or magnitude for 
the reminder; since proofs demonstrate that this asymptotic notation 
does not in fact hide significant coefficients, we believe this 
representation is faithful for the purpose of finite-length analysis, 
as well.

The implications of \cref{lem:torn-lin-red} are more clearly stated in 
the next corollary.
\begin{corollary}\label{cor:torn-asym-rate}
Let $\cC$ be any $(\lmin,\lmax)$-multistrand torn-paper code. 
Assuming $\log(k)=o(n)$, if $\lmin = (a+o_{n k}(1)) \log(n k)$, for 
some $a\geq 1$, then $R(\cC)\leq 1-\frac{1}{a} + o_{n k}(1)$.
\end{corollary}
\begin{IEEEproof}
From \cref{eq:trivialbound} and \cref{lem:torn-lin-red} we have 
\begin{IEEEeqnarray*}{+rCl+x*}
    R(\cC) &\leq&  
    \frac{\log\abs*{\mathset*{\cT_{\lmin}(S)}{S\in\cX_{n,k}}}}{\log\abs*{\cX_{n,k}}} \\
    &\leq& 1 - \frac{n k}{\log\abs*{\cX_{n,k}}} \parenv*{\frac{1}{a} - a \frac{\log(k)}{n} - O\parenv*{\frac{\log\log(n k)}{\log(n k)}}}, 
\end{IEEEeqnarray*}
which, together with $\log\abs*{\cX_{n,k}}\leq n k$, justifies the 
claim.
\end{IEEEproof}

\section{Constructions of Torn-paper Codes}\label{sec:torn-single}

In this section we study constructions of torn-paper codes, in context 
of the bound of \cref{cor:torn-asym-rate}.

\subsection{Related works: pilot-based construction}

An explicit and efficient coding scheme was presented 
in~\cite{ShoVah21} for the probabilistic torn-paper channel. Therein, 
it was argued that an indexing approach to coding is challenging due 
to the a priori unknown locations of segmentation by the channel, 
hence this construction relied on interleaving a \emph{pilot} (or 
\emph{phase-detection sequence}). We describe this scheme below to 
study its performance in the adversarial channel.

\addtocounter{cnstr}{15}
\begin{construction}\cite[Sec.~VII]{ShoVah21}\label{cnst:pilot}
Fix an integer $m>1$. Let $n$ be a multiple of~$m$, to be determined 
later, and $s$ an integer satisfying $s\geq \log(n/m)$. Let 
$\bfp\in \Sigma^{n/m}$ be any $(n/m)$-segment of a de~Bruijn 
sequence~\cite{deBvAE51} of order~$s$, which we refer to as the 
\emph{pilot}.

For $\bfx,\bfy\in \Sigma^{n/m}$, denote $\bfx\per{s} \bfy$ if $\bfx,
\bfy$ have no common $s$-segment, i.e., if for all $i,j\in [n/m-s+1]$ 
it holds that $\bfx^{(i)}\neq \bfy^{(j)}$, where $\bfx^{(i)}$ 
($\bfy^{(j)}$) is the $s$-segment of $\bfx$ (respectively, $\bfy$) at 
location~$i$ (respectively, $j$). Then, we denote $\cO_\bfp\deq 
\mathset*{\bfc\in\Sigma^{n/m}}{\bfc\per{s}\bfp}$.

For any code~$\cC\subseteq \Sigma^{n/m}$, we construct a code 
$\code{pilot}\subseteq \Sigma^n$ as follows: for every choice of $m-1$ 
elements $\parenv*{\bfc_j}_{j\in [m-1]}\subseteq \cC\cap \cO_\bfp$ 
(allowing for repetition), we interleave a single symbol from each 
$\bfp, \bfc_0, \bfc_1, \ldots, \bfc_{m-2}$, in order, to construct a 
codeword $\bfc \in \code{pilot}$.
\end{construction}
\setcounter{cnstr}{0}

\begin{example}\label{exm:pilot}
Let $q=2, m=2, n=12, s=3$. We choose $00010111$ as the binary de~Bruijn 
sequence of order $s$, and let $\bfp\deq 000101$ be its $(n/m)$-prefix. 
Then,
\begin{align*}
    \cO_\bfp = \big\lbrace
    &011100, 011110, \\
    &011111, 111100, \\
    &111110, 111111\big\rbrace.
\end{align*}
Letting $\cC\deq \Sigma^5$, and for any choice of $m-1=1$ element of 
$\cC\cap \cP_\bfp = \cO_\bfp$, we interleave $\bfp$ with that element 
to derive the code 
\begin{align*}
    \code{pilot} = \big\lbrace
    &000101110010, 000101110110, \\
    &000101110111, 010101110010, \\
    &010101110110, 010101110111\big\rbrace.
\end{align*}
\end{example}

\begin{lemma}\cite[Sec.~VII-B]{ShoVah21}\label{lem:pilot-valid}
For all $s\geq \log(n/m)$ it holds that $\code{pilot}$ is an $(ms,
\lmax)$-single strand torn-paper code, for any $\lmax\geq ms$.
\end{lemma}
\begin{IEEEproof}
We replicate the proof for completeness. Observe that every 
$(m s)$-segment~$\bfu$ of $\bfc\in \code{pilot}$ contains $s$ 
consecutive symbols from each $\bfp, \bfc_0, \bfc_1, \ldots, 
\bfc_{m-2}$; since $\bfc_j\per{s}\bfp$ for every $j\in [m-1]$, the 
$s$-segment of $\bfp$ can be uniquely identified. Since $\bfp$ is a 
segment of a de~Bruijn sequence of order~$s$, the location in $\bfp$ 
of the observed segment can be deduced, and hence the location of 
$\bfu$ in $\bfc$ can readily be obtained.
\end{IEEEproof}

\begin{example}\label{exm:pilot-torn}
Continuing \cref{exm:pilot}, assume $010101110110\in \code{pilot}$ is 
passed through an adversarial torn-paper channel with $\lmin = m s = 6$ 
and, say, $\lmax = 8$. The received segments are 
\begin{align*}
    010101, 110110.
\end{align*}
Taking the first segment, we decompose the two interleaved strings 
\begin{align*}
    \bar{\bfc}_0 = 000, \bar{\bfc}_1 = 111;
\end{align*}
we identify $\bar{\bfc}_0$ as 
\emph{the} $s$-substring of $\bfp$ at location~$0$, implying that 
$\bar{\bfc}_1$ is the substring of $\bfc_0$ at location~$0$. Similarly, 
we decompose the second segment into 
\begin{align*}
    \tilde{\bfc}_0 = 101, \tilde{\bfc}_1 = 110;
\end{align*} 
since $101$ is the $s$-substring of $\bfp$ at 
location~$3$, we also have that $\tilde{\bfc}_1$ is the substring of 
$\bfc_0$ at location~$3$, i.e., $\bfc_0 = 111110\in \cO_\bfp$, 
confirming $010101110110\in \code{pilot}$ was the transmitted sequence.
\end{example}

For the probabilistic channel studied in \cite{ShoVah21}, $\cC$ 
in~\cref{cnst:pilot} was chosen to be an error-correcting code. Note 
from the proof of \cref{lem:pilot-valid} that in our chosen 
adversarial setting, this is redundant; that element of the 
construction is preserved in our presentation to support the 
discussion in~\cref{sec:conc} regarding alternate models. 

Next, we turn to find the achievable rates of~\cref{cnst:pilot}.

\begin{corollary}\label{cor:pilot-rate}
$R(\code{pilot}) = \parenv*{1-\frac{1}{m}} R(\cC\cap \cO_\bfp)$.
\end{corollary}
\begin{IEEEproof}
Observe that $\abs*{\code{pilot}} = \abs*{\cC\cap \cO_\bfp}^{m-1} = 
q^{n \parenv*{1-\frac{1}{m}} \log(\abs*{\cC\cap \cO_\bfp}) / 
\frac{n}{m}} = q^{n \parenv*{1-\frac{1}{m}} R(\cC\cap \cO_\bfp)}$.
\end{IEEEproof}

The following lemma was implied by~\cite[Sec.~VII-A]{ShoVah21}.

\begin{lemma}
For all $\cC\subseteq\Sigma^{n/m}$ there exists $\bfz\in\Sigma^{n/m}$ 
such that 
\begin{align*}
	R\parenv*{(\bfz+\cC)\cap \cO_\bfp}\geq R(\cC) - (1 - R(\cO_\bfp)),
\end{align*}
where $\bfz+\cC\deq \mathset*{\bfz+\bfc}{\bfc\in \cC}$.
\end{lemma}
\begin{IEEEproof}
Observe that 
\begin{align*}
	\sum_{\bfz\in \Sigma^{n/m}} \abs*{(\bfz+\cC)\cap \cO_\bfp} 
	&= \sum_{\bfz\in \Sigma^{n/m}} 
	\sum_{\substack{\bfc_1\in \cC \\ \bfc_2\in \cO_\bfp}} 
	\1_{\bfz + \bfc_1 = \bfc_2} \\
	&= \sum_{\substack{\bfc_1\in \cC \\ \bfc_2\in \cO_\bfp}} 
	\sum_{\bfz\in \Sigma^{n/m}} \1_{\bfz = \bfc_2 - \bfc_1} \\
	&= \sum_{\substack{\bfc\in \cC \\	\bfo\in \cO_\bfp}} 1
	= \abs*{\cC}\cdot \abs*{\cO_\bfp}.
\end{align*}
It follows from the pigeonhole principle that there exists $\bfz\in 
\Sigma^{n/m}$ such that $\abs*{(\bfz+\cC)\cap \cO_\bfp}\geq q^{-n/m} 
\abs*{\cC}\cdot \abs*{\cO_\bfp}$, which concludes the proof.
\end{IEEEproof}

In the rest of the section, it remains to analyze what values of~$s$ 
assure that $1-R(\cO_\bfp) = o_n(1)$; we also discuss the implications 
of these available choices.

\begin{lemma}\cite[Sec.~VII-A]{ShoVah21}\label{lem:pilot-union}
If $s\deq \ceilenv*{(2+\delta) \log(n/m)}$ for some $\delta>0$, then, 
using $\cC\deq \Sigma^{n/m}$ in \cref{cnst:pilot}, 
\begin{align*}
    R(\code{pilot}) 
    &\geq 1 - \frac{1}{m} - \frac{m-1}{n}\cdot\frac{1}{(n/m)^\delta-1} \\
    &= 1 - \frac{1}{m} - o_n(1).
\end{align*}
\end{lemma}
\begin{IEEEproof}
Again, we replicate the proof here. Denote for a uniformly chosen 
$\bfc\in \Sigma^{n/m}$ the event $A_{i,j}$ that $\bfc^{(i)} = 
\bfp^{(j)}$. Clearly $\Pr(A_{i,j})=q^{-s}$; using the union bound, 
$\Pr(\bfc\per{s} \bfp)\geq 1 - (n/m)^2 q^{-s}\geq 1-(n/m)^{-\delta}$, 
i.e., 
\begin{align*}
	\abs*{\cO_\bfp} \geq q^{n/m} \parenv*{1-(n/m)^{-\delta}}.
\end{align*}
It follows from \cref{cor:pilot-rate} that 
\begin{IEEEeqnarray*}{+rCl+x*}
	R(\code{pilot}) &=& \parenv*{1-\frac{1}{m}} R(\cO_\bfp) \\
	&\geq& \parenv*{1-\frac{1}{m}} \parenv*{1 + \frac{m}{n} 
	\log\parenv*{1-(n/m)^{-\delta}}} \\
	&\geq& 1 - \frac{1}{m} - 
	\frac{(m-1)(n/m)^{-\delta}}{n (1-(n/m)^{-\delta})}. 
	\\[-\normalbaselineskip] &&&\IEEEQEDhere
\end{IEEEeqnarray*}
\end{IEEEproof}

Unfortunately, \cref{lem:pilot-union} doesn't match the upper bound of 
\cref{cor:torn-asym-rate}; asymptotically, it produces rate $1-
\frac{2+\delta}{a}$, where $a\deq \frac{m s}{\log(n)}$. Further, the 
construction may only be applied when $a$ is (approximately) an 
\emph{even integer} $\geq 4$. 
The former can be remedied by replacing the union bound in the 
analysis of~\cite[Sec.~VII-A]{ShoVah21} with the Lov\'{a}sz local 
lemma~\cite{Spe77} (similarly to techniques used independently in 
\cite{YehPol21} and \cite{EliGabMedYaa21}), as follows.

\begin{lemma}\label{lem:pilot-lovasz}
Let $s\deq \ceilenv*{\log(n/m)+\log\log(n/m)+\log(3e)}$. Then, using 
$\cC\deq \Sigma^{n/m}$ in \cref{cnst:pilot}, 
\begin{align*}
    R(\code{pilot}) 
    &\geq \parenv*{1 - \frac{1}{m}}\cdot \parenv*{1 - \frac{\log(e)}{2\log(n/m)}} \\
    &= 1 - \frac{1}{m} - O\parenv*{\frac{1}{\log(n)}}.
\end{align*}
\end{lemma}
\begin{IEEEproof}
Denote for a uniformly chosen $\bfc\in \Sigma^{n/m}$ the event 
$A_{i,j}$ that $\bfc^{(i)} = \bfp^{(j)}$. Clearly $p\deq \Pr(A_{i,j}) 
= q^{-s}$, and $A_{i,j}$ is jointly independent of 
$\mathset*{A_{i',j'}}{\abs*{i-i'}\geq s}$, 
i.e., all except $(n/m-s) (2s-1) - 1\leq 2s n/m - 1$ distinct events.

For sufficiently large $n$, observe that 
\begin{align*}
	s q^{-s} &\leq 
	\frac{\log(n/m)+\log\log(n/m)+\log(3e)}{(n/m) \log(n/m) 3e} \\
	&= \frac{m}{2e n}\cdot 
	\frac{2}{3} \parenv*{1 + \frac{\log\log(n/m)+\log(3e)}{\log(n/m)}} 
	< \frac{m}{2e n},
\end{align*}
where the first inequality is justified by $(s+r) q^{-(s+r)}\leq 
s q^{-s}$ for $r\geq 0$ and $s\geq \log(e)$. 
Rearranging, we have $m\geq 2e p s n$. Therefore, letting $x\deq 
\frac{e p}{1+e p}$ (hence, $\frac{x}{1-x} = e p$), and recalling for 
all $x\in (0,1)$ that $1-x\geq \exp(\frac{-x}{1-x})$, we have 
\begin{align*}
    x (1-x)^{2s n/m - 1} 
    &= e p (1-x)^{2s n/m} \\
    &\geq p \exp\parenv[\bigg]{1 - 2e p s n/m} > p.
\end{align*}
It therefore follows from the local lemma that 
\begin{align*}
	\Pr(\bfc\per{s} \bfp) 
	&\geq (1-x)^{(n/m)^2} \geq \exp\parenv*{-ep (n/m)^2} \\
	&= e^{-e (n/m)^2 q^{-s}} 
	\geq e^{-n/2s m}, 
\end{align*}
where again we used the fact that $m\geq 2e p s n$. That is, 
$\abs*{\cO_\bfp}\geq q^{n/m} e^{-n/2s m} = 
\parenv*{q e^{-1/2s}}^{n/m}$, and 
\begin{align*}
	R(\cO_\bfp) 
	&\geq 1 - \frac{\log(e)}{2s}. 
\end{align*}
Hence, \cref{cor:pilot-rate} concludes the proof.
\end{IEEEproof}

Based on \cref{lem:pilot-lovasz}, \cref{cnst:pilot} achieves $1 - 
\frac{1}{a} - o_n(1)$ rate, where $a\deq \frac{m s}{\log(n)}$, 
asymptotically matching the bound of \cref{cor:torn-asym-rate}. It 
also expands the values of $a$ for which the construction may be 
applied; however, unfortunately $a$ is still restricted to be 
(approximately) an \emph{integer} $\geq 2$. 
Moreover, encoding $\code{pilot}(n)$ involves a choice 
of~$\bfp$, and the authors are not aware of a straightforward way to 
make this choice while optimizing $R(\cO_\bfp)$; it further requires 
encoding into (potentially, a sub-code of) $\cO_\bfp$, which is also, 
to the best of our knowledge, not readily done in an efficient manner. 
To bridge that gap, we present in the next section a construction based 
on an indexing approach, which can be applied for any $a>1$, 
asymptotically matching \cref{cor:torn-asym-rate} for all choices.

\subsection{Index-based construction}

In this section, an index-based construction of single-strand 
torn-paper codes is presented and is then extended for multiple 
strands.

It is assumed from here on out that $\lmin = \ceilenv*{a \log(n)}$, 
for some $a>1$ which is fixed throughout this section. 
We propose the following construction of length-$n$ $(\lmin, 
\lmax)$-single strand torn-paper codes. 
The construction is based on the following components.

\begin{figure}[t]{}%
\centering
\psfrag{parity symbol}{parity symbol}
\includegraphics[width=0.95\columnwidth]{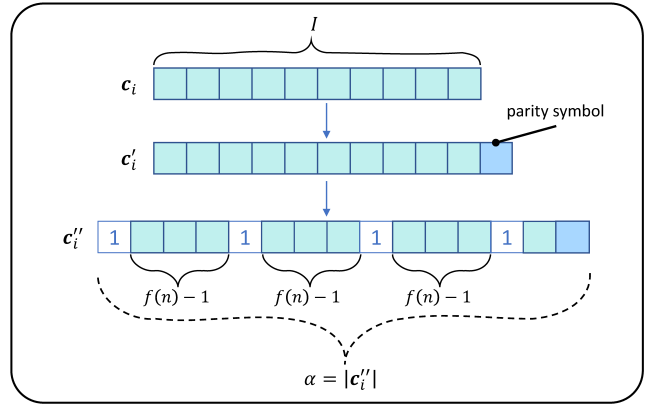}%
\caption{Index generation (best viewed in online colored version) \label{fig:index_torn}}
\end{figure}

\begin{definition}
For an integer $I$, let $\parenv*{\bfc_i}_{i\in[q^I]}$, $\bfc_i\in 
\Sigma^I$ be codewords of a $q$-ary Gray code, in order. 
Denote by $\bfc'_i$ the concatenation of $\bfc_i$ with a single parity 
symbol (i.e., the sum of the entries in $\bfc'_i$ is zero). 
Further, denote by $\bfc''_i$ the result of inserting `$1$'s into 
$\bfc'_i$ at every location divisible by $f(n)$ (since the locations of 
substrings start with $0$, the first bit of $\bfc''_i$ is always 
`$1$'). The process is illustrated in \cref{fig:index_torn}. Note that 
$\alpha\deq \abs*{\bfc''_i} = \ceilenv*{\frac{f(n)}{f(n)-1} (I+1)}$ for 
all $i\in [q^I]$. 
We refer to $\bfc_i$ (or simply~$i$) as an \emph{index} in the 
construction and to $\bfc''_i$ as an \emph{encoded index}.
\end{definition}

This index structure is motivated by the property indicated in the 
following lemma.
\begin{lemma}\label{lem:index-concat}
Let $\bfc$ be an $\alpha$-substring of $\bfc''_i\circ \bfc''_{i+1}$, 
for some $i\in [q^I-1]$. Then $i$ can uniquely be recovered 
from~$\bfc$.
\end{lemma}
\begin{IEEEproof}
Since $\bfc''_i$ and $\bfc''_{i+1}$ differ only at the parity symbol 
and one additional coordinate (which corresponds to the only position 
where $\bfc_i$ and $\bfc_{i+1}$ differ), $\bfc$ is either $\bfc''_i$ 
or a copy of $\bfc''_{i+1}$ with an erroneous parity symbol. 
To obtain~$i$ it suffices to distinguish these two cases, which may be 
done by calculating the parity symbol of $\bfc''$; If the parity 
symbol is correct then $i$~equals to the decoding of 
$\bfc$ (with the Gray-code decoder), and otherwise $i$~equals to the 
decoding of $\bfc$ minus one.
\end{IEEEproof}

\begin{definition}
Let $f,N$ be integers. 
The \emph{Run-length limited (RLL) encoder} $E_N^{RLL}$ receives 
strings of length~$m(N)$ and returns strings of length~$N$ that do not 
contain zero runs of length $f$. Constructions of such encoders can be 
taken from \cite{LevYaa19} or \cite[Lem.~4]{YehBarMarYaa23}.
\end{definition}

\begin{figure}[t]{}%
\centering
\psfrag{parity symbol}{parity symbol}
\includegraphics[width=1.00\columnwidth]{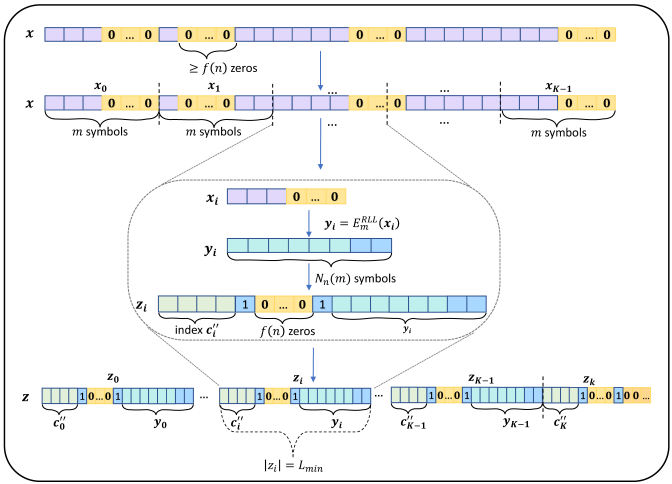}%
\caption{Illustration of~\cref{alg:torn-gray-encode} (best viewed in online colored version) 
\label{fig:const_torn}}
\end{figure}

\begin{construction}\label{cnst:torn-gray}
The main idea of the construction is that every codeword should 
constitute a concatenation of length-$\lmin$ segments with the 
following structure: an index, followed by a marker, then encoded data.
Let $f(n)$ be any integer-valued function satisfying $f(n)=\omega(1)$ 
and $f(n)=o(\log(n))$ (see \cref{thm:torn-gray-red} for a choice 
optimizing the redundancy of this construction). Further assume $n\geq 
\lmin\geq \alpha+f(n)+2$. 
Let $I\deq\ceilenv*{\log(n/\lmin)}$, $K\deq \floorenv*{n/\lmin} - 1$ 
and $N\deq \lmin - \alpha - f(n) - 2$. 
The constructed $(\lmin,\lmax)$-single strand torn-paper code, denoted 
by $\code{cnst:torn-gray}(n)$, is defined by the encoder 
${\enc[cnst:torn-gray]:\Sigma^{K m(N)}\to \Sigma^n}$ 
in~\cref{alg:torn-gray-encode}, and illustrated 
in~\cref{fig:const_torn}.
\end{construction}
\begin{algorithm}[t]
\caption{Encoder for \cref{cnst:torn-gray}}
\label{alg:torn-gray-encode}
\SetKwInput{KwInput}{Input}                
\SetKwInput{KwOutput}{Output}  
\SetAlgoLined
\KwInput{ $\bfx =(x_0, x_1, \ldots, x_{K m(N)-1})\in \Sigma^{K m(N)}$}  
\KwOutput{$\enc[cnst:torn-gray](\bfx)$}
\For{$i\leftarrow 0$ to $K-1$}{
    $\bfx_i\leftarrow (x_{i m(N)}, x_{i m(N)+1},\ldots ,x_{(i+1) m(N)-1})$ \tcp{$\abs*{\bfx_i}=m(N)$}
    $\bfy_i\leftarrow E_N^{RLL}(\bfx_i)$ \tcp{$\bfy_i$ contains no zero runs of length~$f(n)$}
    $\bfz_i \leftarrow \bfc''_i\circ 1 0^{f(n)} 1\circ \bfy_i$ \tcp{$\abs*{\bfz_i}=\lmin$}
}
$\bfz_K \leftarrow  \bfc''_K \circ 1 0^{f(n)} 1 0^{N}$ \tcp{$\abs*{\bfz_K}=\lmin$}
$\bfz \leftarrow \bfz_0\circ \bfz_1\circ \cdots\circ \bfz_K \circ  0^{n\bmod\lmin}$ \tcp{$\abs*{\bfz} = n$} 
\Return $\bfz$
\end{algorithm}

In the rest of the paper, we call the strings $\bfx_i$ (respectively, 
$\bfy_i$) in the constructions an \emph{information block} 
(\emph{encoded block}); the strings $1 0^{f(n)} 1$ are called 
\emph{markers}; finally, a string~$\bfz_i$ will simply be referred 
to as a segment of $\bfz$. 
Note that the last segment~$\bfz_K$ of~$\bfz$ deliberately 
does not contain data, to account for the possibility that a part 
of~$\bfz_K\circ 0^{n\bmod\lmin}$ might be partitioned by an adversarial 
channel in such a way that it does not contain, at its suffix, a prefix 
of an index. 
We observe that once the encoded blocks 
$\bfy_i$'s are obtained, encoding (including the generation of the 
Gray code) then requires a number of operations linear in~$n$. 
By~\cite{LevYaa19,YehBarMarYaa23}, encoding each $\bfx_i$ into $\bfy_i$ 
may also be achieved with a linear number of operations. Hence, the 
complexity of \cref{cnst:torn-gray} is linear with~$n$. 

\begin{example}\label{exm:torn-gray-encode}
We demonstrate the operation of $\enc[cnst:torn-gray]$. 
Let $q = 2, n = 45, \lmin = 14, f(n) = 2$. 
For index generation, we utilize the binary Gray code $\parenv*{00, 01, 
11, 10}$, whose encoded indices are (in order) 
\begin{align*}
    \parenv*{101010, 101111, 111110, 111011}
\end{align*}
(observe $\alpha=6$). 
Let $N=\lmin-f(n)-2-\alpha=4$, and observe an encoder~$E_N^{RLL}$ 
exists with $m(N)=3$, defined by the lexicographic ordering of the 
$f(n)$-run-length-limited sequences of length~$N$ 
\begin{align*}
    \big\lbrace &0101, 0110, 0111, 1010, \\*
    &1011, 1101, 1110, 1111\big\rbrace.
\end{align*}
Noting that $K = 2$, we demonstrate, e.g., the encoding of the 
information sequence $001110$. Observe, $\bfx_0 = 001, \bfx_1 = 110$, 
hence $\bfy_0 = 0110, \bfy_1 = 1110$. We then have 
\begin{align*}
    \bfz_0 &= 101010\ 1001\ 0110 \\
    \bfz_1 &= 101111\ 1001\ 1110 \\
    \bfz_2 &= 111110\ 1001\ 0000, 
\end{align*}
and $\bfz = \bfz_0\circ \bfz_1\circ \bfz_2\circ 000$.
\end{example}

Next, it is shown that the constructed code $\code{cnst:torn-gray}(n)$ 
is an $(\lmin,\lmax)$-single strand torn-paper code.
\begin{theorem}\label{thm:torn-codes}
For all $\lmax\geq\lmin$, $\code{cnst:torn-gray}(n)$ is an $(\lmin,
\lmax)$-single strand torn-paper code with a linear-run-time decoder.
\end{theorem}

The proof of \cref{thm:torn-codes} is carried by presenting an 
explicit decoder to $\code{cnst:torn-gray}(n)$ as follows. Let $\bfz 
\in \code{cnst:torn-gray}(n)$ and let $\bfz = \bfu_0\circ \bfu_1\circ 
\cdots\circ \bfu_{s-1}$ so that $\multbrace*{\bfu_0,\bfu_1,
\ldots, \bfu_{s-1}}$ is an $(\lmin,\lmax)$-segmentation of $\bfz$. 
The main task of the decoding algorithm is to successfully retrieve 
the location within $\bfz$ of each of the $s$~segments of the $(\lmin, 
\lmax)$-segmentation. 
For every segment $\bfu_j$, $j\in[s]$, the decoder first finds the 
location~$i$ such that the first (maybe partial) occurrence of an 
encoded index in the segment $\bfu_j$ is of~$\bfc_i''$ (see below for 
a proof that this is possible). Given $i$ and the location of 
$\bfc_i''$ in $\bfu_j$, the location of the segment $\bfu_j$ within 
$\bfz$ can be calculated. Then, according to the location in $\bfz$ 
for each segment in the $(\lmin, \lmax)$-segmentation, one can simply 
concatenate the segments in the correct order to obtain the 
codeword~$\bfz$. Finally, by removing the markers and the encoded 
indices and applying the RLL decoder for each of the strings 
$\bfy_i$'s, the information string~$\bfx$ is retrieved. 

Consider the case where a segment $\bfu$ is a proper substring of the 
suffix of $\bfz$ of length $(n\bmod\lmin)+N+f(n)$, i.e., $\bfz_K 
0^{n\bmod\lmin}$ (note that this does not imply that $\bfu$ is itself 
a suffix of $\bfz$). Then, $\bfu$ does not intersect $\bfy_i$ for any 
$i\in[K]$, and may safely be discarded. We see next that these cases 
may be identified efficiently.
\begin{lemma}\label{lem:suffix-iden}
Let $\bfz\in \code{cnst:torn-gray}(n)$ and let $\bfu$ be a proper 
substring of $\bfz_K 0^{n\bmod\lmin}$. If $n$ is sufficiently large 
(specifically, if $(a-1) \ceilenv*{\log(n)} > 2 f(n) + 1$), then this 
fact can efficiently be identified.
\end{lemma}
\begin{IEEEproof}
Observe that either $\abs*{\bfu}<\lmin$ or $\bfu$ contains a suffix of 
`$0$'s of length at least 
\begin{align*}
	\lmin-\alpha-f(n)-1 &\geq (a-1) \ceilenv*{\log(n)}-f(n)-1,
\end{align*}
i.e., longer than $f(n)$, which can easily be identified. 
\end{IEEEproof}

By \cref{lem:suffix-iden}, it is sufficient to retrieve the location 
of any segment which is not a substring of the suffix of length 
$(n\bmod\lmin)+N+f(n)$ of $\bfz$. For any such $\bfu$, the 
calculation of the index~$i$ such that $\bfc_i''$ is the first 
(perhaps partial) occurrence of an encoded index within $\bfu$, is 
given in~\cref{alg:decode-index-lmin}.  

Any $L$-segment $\bfu$ of $\bfz\in \code{cnst:torn-gray}(n)$, such 
that $L\geq\lmin$, contains at least part of one of the encoded 
indices $\bfc_i''$. If $\bfc''_i$ is the first encoded index to 
intersect $\bfu$, we denote by $\ind(\bfu)\deq i$ the \emph{index of 
$\bfu$}. Note that this index does not depend on the information that 
was encoded in the construction, but rather, only on the location of 
$\bfu$ in $\bfz$. 
\cref{alg:decode-index-lmin} ensures that it is possible to determine 
the index of every $L$-segment $\bfu$ of $\bfz$, where $L\geq\lmin$.

\begin{algorithm}[t]
\caption{Index retrieval from a segment}
\label{alg:decode-index-lmin}
\SetKwInput{KwInput}{Input}                
\SetKwInput{KwOutput}{Output}  
\SetAlgoLined
\KwInput{An $L$-segment $\bfu$ of a codeword of 
$\code{cnst:torn-gray}(n)$, where $L\geq \lmin$.}  
\KwOutput{The index of $\bfu$ within $\bfz$, $\ind(\bfu)$}
$\bfu' \leftarrow$ the $\lmin$-length prefix of $\bfu$\\ 
 $j\leftarrow$ the starting index of the unique occurrence of $10^{f(n)}1$ within $\bfu'$; if none exists, of the cyclic occurrence \\  
 $\bfc''\leftarrow$ the (cyclic) $\alpha$-substring of 
    $\bfu$ strictly preceding $j$\\ 
$\bfc'\leftarrow$ the non-padded subsequence of $\bfc''$ \\
$\bfc\leftarrow$ the $I$-prefix of $\bfc'$ \\
$\ind \leftarrow$ the index of $\bfc$ in the Gray code \\

    \If{the last symbol of $\bfc'$ is not the parity of $\bfc$}
    {
         $\ind\leftarrow \ind -1$
    } 
    \Return $\ind$
\end{algorithm}

The correctness of \cref{alg:decode-index-lmin} follows from 
the next lemma. 
\begin{lemma}\label{lem:torn-gray}
Let $\bfz\in \code{cnst:torn-gray}(n)$, $L\geq \lmin$, and let $\bfu$ 
be an $L$-segment of $\bfz$ which is not a substring of the suffix of 
length $(n\bmod\lmin)+N+f(n)$ of $\bfz$. Then, 
\cref{alg:decode-index-lmin} successfully returns the 
index~$\ind(\bfu)$ of $\bfu$. 
\end{lemma}
\begin{IEEEproof}
Let $\bfu$ be a substring of $\bfz$ and w.l.o.g. assume that 
$\abs*{\bfu}=\lmin$. From the RLL encoding of the strings $\bfx_i$'s, 
observe that $\bfu$ does not contain any occurrences of $10^{f(n)}1$ 
except those explicitly added to the encoded indices by 
\cref{cnst:torn-gray}. 
Since $\abs*{\bfz_j}=\lmin$ for all $j$, either $\bfu$ contains an 
occurrence of $1 0^{f(n)} 1$ or it has a suffix-prefix pair whose 
concatenation is $1 0^{f(n)} 1$ 
(this follows from \cref{cnst:torn-gray} and the assumption that 
$\bfu$ does not begin with a proper suffix of $\bfz_K 
0^{n\bmod\lmin}$). In both cases, we will show that the precise 
location of the (perhaps incomplete) occurrence of $\bfc''_i$ in 
$\bfu$ can be deduced, for some~$i$. 

Let $j$ be the (unique) location in $\bfu$ of the substring $1 0^{f(n)}
1$. If $j\geq \alpha$, then $\bfu$ contains a complete occurrence of 
the encoded index $\bfc''_i$, and so the index $\bfc_i$, and 
therefore~$i$, are readily obtained. 
Otherwise, $j < \alpha$ and let $\bfc''$ be the cyclic 
$\alpha$-substring of $\bfu$ strictly preceding the substring $1 
0^{f(n)}1$ which starts at location $\lmin-(\alpha-j)$. 
The substring $\bfc''$ is obtained by the concatenation of the 
$(\alpha-j)$-suffix of $\bfu$ with the $j$-prefix of $\bfu$. 
The proof is now concluded by \cref{lem:index-concat}.
\end{IEEEproof}

We remark that the described procedure operates in run-time which is 
linear in the substring length. In addition, if $\bfz$ can be 
reconstructed from its non-overlapping substrings, then the strings 
$\bfy_i$'s are readily obtained, and $\bfx$ may be decoded (again, see 
\cite{LevYaa19,YehBarMarYaa23}). These algorithms also require a linear 
number of operations. This completes the proof 
of~\cref{thm:torn-codes}.

\begin{example}\label{exm:torn-gray-decode}
We return to \cref{exm:torn-gray-encode}, to demonstrate the operation 
of~\cref{thm:torn-codes}. Recall, for $q = 2, n = 45, \lmin = 14, f(n) 
= 2$, that we have constructed the following codeword 
\begin{align*}
    \bfz &= 101010100101101011111001111011111010010000000.
\end{align*}
Suppose that we receive the following $(14,20)$-segmentation of~$\bfz$:
\begin{align*}
    \multbrace*{10101010010110101,1111001111011111,010010000000}.
\end{align*}
Note since $\abs*{010010000000}=12<\lmin$, it might readily be inferred 
that it is the suffix of~$\bfz$. We therefore only need identify the 
locations of the other two segments.
\begin{itemize}
\item 
The segment $101010\ 1001\ 0110101$ contains the marker $10^{f(n)}1 = 
1001$, and we therefore conclude that $101010$ (given $\alpha=6$) is an 
encoded-index, which as we recall corresponds to the Gray-code 
element~$c_0 = 00$. It follows that $\bfy_0 = 0110$, and $101$ is a 
prefix of~$\bfz_1$ (observe that the following segment of~$\bfz_1$ 
could not have been immediately identify, if more segments were 
received).

\item 
Next, the segment $111\ 1001\ 111011111$ also contains a marker, 
implying that $111$ is the suffix of~$c''_i$, and $111$ the prefix 
of~$c''_{i+1}$. Concatenating, we have the index~$\bfc'' = 111111$ and 
$\bfc' = 11\ 1$, which is an instance of~$\bfc_2$ containing an 
erroneous parity symbol. Hence we deduce $i=1$, and $\bfy_1 = 1110$.
\end{itemize}
Together, the decoding $\bfx_0 = 001$ and $\bfx_1 = 110$ may now be 
performed, reconstructing the original information sequence.
\end{example}

Lastly the redundancy of~\cref{cnst:torn-gray} is analyzed.
\begin{theorem}\label{thm:torn-gray-red}
Using the RLL encoders of \cite{LevYaa19,YehBarMarYaa23} in 
\cref{cnst:torn-gray}, it holds that 
\begin{align*}
    \red(\code{cnst:torn-gray}(n)) &\leq 
    \frac{n}{a} \bigg(1 + \frac{f(n)}{\log(n)} + \frac{1}{f(n)-1} \>+ \\*
    &\hphantom{\leq \frac{n}{a} \bigg(} \frac{9 + 
    2/(f(n)-1)}{\log(n)} + \frac{4a}{q^{f(n)}} + \frac{2 a^2 + 2}{n}\bigg) \\
    &= \frac{n}{a} \parenv*{1 + 
	(1+o(1)) \parenv*{\frac{f(n)}{\log(n)} + \frac{1}{f(n)}}}.
\end{align*}
In particular, the redundancy is optimized for $f(n) = (1+o(1)) 
\sqrt{\log(n)}$, i.e.,
\begin{align*}
    \red(\code{cnst:torn-gray}(n)) &\leq \frac{n}{a} \parenv*{1 + 
    \frac{2+o(1)}{\sqrt{\log(n)}}}.
\end{align*}
\end{theorem}
\begin{IEEEproof}
From \cref{cnst:torn-gray}, observe that 
$\red(\code{cnst:torn-gray}(n)) = (n\bmod \lmin) + \lmin + 
K(\lmin-m(N))$ 
and 
\begin{align*}
    \lmin-N &= \alpha+f(n)+2 \\
    &= \ceilenv*{\frac{f(n)}{f(n)-1} (I+1)} + f(n) + 2 \\
    &\leq \frac{f(n)}{f(n)-1} (I+1) + f(n) + 3 \\
    &\leq \frac{f(n)}{f(n)-1} \parenv*{\log(n/\lmin)+2} + f(n) + 3 \\
    &\leq \log(n) + f(n) + \frac{\log(n)}{f(n)-1} + 5 + \frac{2}{f(n)-1}.
\end{align*}

Further, by \cite[Lem.~4]{YehBarMarYaa23} one may efficiently encode 
$\bfx\mapsto \bfy$ such that 
$
	N - m(N) \leq \ceilenv[\big]{\frac{q}{q-2}\cdot 
	\frac{N}{q^{f(n)}}}
$
(For $q=2$ \cite[Sec.~III]{LevYaa19} showed $N - m(N) \leq 
2\ceilenv*{N/q^{f(n)-1}}$), and we shall use the overly zealous upper 
bound $N - m(N) \leq \frac{4 N}{q^{f(n)}} + 2\leq \frac{4 a 
\log(n)}{q^{f(n)}} + \frac{4}{q} + 2\leq \frac{4 a \log(n)}{q^{f(n)}} + 
4$.

Finally, we get that 
\begin{IEEEeqnarray*}{+rCl+x*}
    \red(\code{cnst:torn-gray}(n)) &=& K (\lmin - m(N)) + \lmin + 
    (n\bmod\lmin) \\
    &\leq& \frac{n}{a \log(n)} \bigg(\log(n) + f(n) 
    + \frac{\log(n)}{f(n)-1} \>+ \\*
    && \hphantom{\frac{n}{a \log(n)} \bigg(} 9 + \frac{2}{f(n)-1} 
    + \frac{4 a \log(n)}{q^{f(n)}}\bigg) \>+ \\*
    && \IEEEeqnarraymulticol{1}{r}{2 \lmin} \\
    &\leq& \frac{n}{a} \bigg(1 + \frac{f(n)}{\log(n)} + \frac{1}{f(n)-1} \>+ \\*
    && \hphantom{\frac{n}{a} \bigg(} \frac{9 + 2/(f(n)-1)}{\log(n)} + \frac{4a}{q^{f(n)}} + \frac{2 a^2 + 2}{n}\bigg), 
\end{IEEEeqnarray*}
which completes the proof of the first part. The second part follows 
by substitution of $f(n) = (1+o(1)) \sqrt{\log(n)}$ into the former. 
\end{IEEEproof}

By~\cref{thm:torn-gray-red,cor:torn-asym-rate}, efficient encoding and 
decoding is possible at asymptotically optimal rates. 
In comparison to \cref{cnst:pilot} (by \cref{lem:pilot-lovasz}), 
\cref{cnst:torn-gray} asymptotically achieves rate $1-\frac{1}{a} - 
O\parenv*{\frac{f(n)}{\log(n)}+\frac{1}{f(n)}}$ instead of $1 - 
\frac{1}{\ceilenv*{a}} - O\parenv*{\frac{1}{\log(n)}}$, for any channel 
parameter $a>1$ (here, the integer value is used since 
\cref{cnst:pilot} must be operated at $m\deq\ceilenv*{a}$ to produce an 
$(\lmin,\lmax)$-torn-paper code). 
For completeness, we also include specific construction parameters for 
several arbitrary choices of $n,\lmin$, and compare resulting rates, in 
\cref{tab:union-rates,tab:lovasz-rates,tab:torn-gray-rates} (all for 
$q=4$). It should however be stressed that, for \cref{cnst:pilot}, the 
choice of~$s,m$ optimizing the resulting rate $R(\code{pilot}(n))\geq 
\parenv*{1-\frac{1}{m}}\cdot R(\cO_\bfp)$ is not straightforward, even 
given the lower bounds of \cref{lem:pilot-union,lem:pilot-lovasz}; 
indeed, $R(\cO_\bfp)$ cannot easily be computed, for an optimal choice 
of~$\bfp$. We rely in our comparison on the lower-bounds of 
\cref{lem:pilot-union,lem:pilot-lovasz} instead; note in particular 
that even for the same choice of~$n,m,s$, i.e., for a specific code, 
these might provide distinct lower-bounds on the rate. As mentioned 
above, even then it is not immediately clear how to efficiently encode 
and decode $\code{pilot}(n)$.
\begin{table*}[t]
	\caption{\cref{cnst:pilot} (\cref{lem:pilot-union}): Specific Parameters ($m,s,R(\code{pilot}(n))$).\label{tab:union-rates}}
	\begin{center}
        \begin{tabular}{c|c|c|c|c|c|c}
            $\lmin\big\backslash n$ & $60$ & $250$ & $4000$ & $60,000$ & $400,000$ & $6,000,000$ \\
            \cline{1-7} 
            10 & $2,5,0.379$ & n/a & n/a & n/a & n/a & n/a \\
            \cline{1-7} 
            50 & $15,3,\mathbf{0.856}$ & $10,5,\mathbf{0.844}$ & $5,10,\mathbf{0.798}$ & $3,16,0.667$ & $2,25,0.5$ & $2,25,0.5$ \\
            \cline{1-7} 
            100 & n/a & $10,10,\mathbf{0.9}$ & $10,10,\mathbf{0.9}$ & $6,16,0.833$ & $5,20,0.8$ & $4,25,0.75$ \\
            \cline{1-7} 
            300 & n/a & n/a & $32,9,\mathbf{0.968}$ & $25,12,\mathbf{0.96}$ & $20,15,\mathbf{0.95}$ & $15,20,\mathbf{0.933}$ \\
            \cline{1-7} 
            1000 & n/a & n/a & $125,8,\mathbf{0.992}$ & $100,10,\mathbf{0.989}$ & $64,15,\mathbf{0.984}$ & $50,20,\mathbf{0.98}$
        \end{tabular}

        \vspace{0.5\baselineskip}
        {(\scriptsize Bold-face indicates \cref{lem:pilot-union} provides highest lower-bound on rate.)}
	\end{center}%
	\caption{\cref{cnst:pilot} (\cref{lem:pilot-lovasz}): Specific Parameters ($m,s,R(\code{pilot}(n))$).\label{tab:lovasz-rates}}
	\begin{center}
        \begin{tabular}{c|c|c|c|c|c|c}
            $\lmin\big\backslash n$ & $60$ & $250$ & $4000$ & $60,000$ & $400,000$ & $6,000,000$ \\
            \cline{1-7} 
            10 & $2,5,\mathbf{0.45}$ & n/a & n/a & n/a & n/a & n/a \\
            \cline{1-7} 
            50 & $15,3,0.778$ & $10,5,0.81$ & $5,10,0.76$ & {\cellcolor[gray]{.85}} $5,10,\mathbf{0.76}$ & {\cellcolor[gray]{.85}} $4,12,\mathbf{0.719}$ & {\cellcolor[gray]{.85}} $3,16,\mathbf{0.646}$ \\
            \cline{1-7} 
            100 & n/a & $10,10,0.855$ & $10,10,0.855$ & {\cellcolor[gray]{.85}} $10,10,\mathbf{0.855}$ & {\cellcolor[gray]{.85}} $8,12,0.839$ & {\cellcolor[gray]{.85}} $6,16,0.807$ \\
            \cline{1-7} 
            300 & n/a & n/a & $25,12,0.92$ & $25,12,0.92$ & $25,12,0.92$ & {\cellcolor[gray]{.85}} $20,15,0.918$ \\
            \cline{1-7} 
            1000 & n/a & n/a & $50,20,0.956$ & $50,20,0.956$ & $50,20,0.956$ & $50,20,0.956$
        \end{tabular}

        \vspace{0.5\baselineskip}
        {(\scriptsize Bold-face indicates \cref{lem:pilot-lovasz} provides highest lower-bound on rate. Background pattern indicates that the choice of $m,s$ is only guaranteed by \cref{lem:pilot-lovasz}.)}
	\end{center}%
	\caption{\cref{cnst:torn-gray} (\cref{thm:torn-gray-red}): Specific Parameters ($f,I,N,K,R(\code{cnst:torn-gray}(n))$).\label{tab:torn-gray-rates}}
	\begin{center}
        \begin{tabular}{c|c|c|c|c|c|c}
            $\lmin\big\backslash n$ & $60$ & $250$ & $4000$ & $60,000$ & $400,000$ & $6,000,000$ \\
            \cline{1-7} 
            10 & n/a & n/a & n/a & n/a & n/a & n/a \\
            \cline{1-7} 
            50 & n/a & $2,2,40,4,0.56$ & $3,4,38,79,0.711$ & $3,6,35,1199,0.659$ & $4,7,34,7999,0.66$ & $4,9,31,119999,0.6$ \\
            \cline{1-7} 
            100 & n/a & $2,1,92,1,0.32$ & $3,3,89,39,0.839$ & $3,5,86,599,0.829$ & $4,6,85,3999,\mathbf{0.84}$ & $4,8,82,59999,\mathbf{0.81}$ \\
            \cline{1-7} 
            300 & n/a & n/a & $3,2,291,12,0.843$ & $3,4,288,199,0.925$ & $4,6,9,285,1332,0.939$ & $4,8,282,19999,0.93$ \\
            \cline{1-7} 
            1000 & n/a & n/a & $3,1,992,3,0.721$ & $3,3,989,59,0.942$ & $4,5,986,399,0.976$ & $4,7,984,5999,0.976$
        \end{tabular}

        \vspace{0.5\baselineskip}
        {(\scriptsize Bold-face indicates \cref{thm:torn-gray-red} provides highest lower-bound on rate.)}
	\end{center}
\end{table*}

Next, we consider the case of $k>1$ and $\log(k)=o(n)$. We know 
from~\cref{cor:torn-asym-rate} that if $\limsup\frac{\lmin}{n k} \leq 
1$ then any family of $(\lmin,\lmax)$-multistrand torn-paper codes 
will only achieve vanishing asymptotic rate; hence we assume  $\lmin = 
\ceilenv*{a \log(n k)}$ for some $a>1$. The following theorem 
summarizes our main results regarding $(\lmin,\lmax)$-multistrand 
torn-paper codes.
\begin{theorem}\label{thm:multistrand}
Take $n,k$ such that $k>1$, $\log(k)=o(n)$, and let $\lmin = 
\ceilenv*{a \log(n k)}$, for $a>1$. There exists a linear run-time (in 
the substrings length, i.e., $nk$) encoder-decoder pair for $(\lmin,
\lmax)$-multistrand torn-paper codes achieving $1 - \frac{1}{a} - 
o_{n k}(1)$ asymptotic rate.
\end{theorem}
\begin{IEEEproof}
\cref{thm:multistrand} is justified by a simple amendment of 
\cref{cnst:torn-gray}. We encode $\bfx\in\Sigma^{kKm}$ into 
$\multset*{\bfz^{(j)}}{j\in [k]}$, where $\abs*{\bfz^{(j)}} 
= n$ for all~$j\in [k]$, as follows. We modify $I\deq 
\ceilenv*{\log\parenv*{k \ceilenv*{n/\lmin}}}$ (recall, also, $\alpha 
\deq \ceilenv[\big]{\frac{f(n)}{f(n)-1} (I+1)}$) and $\lmin = 
\ceilenv*{a \log(n k)}$. We then denote $\bfx = \bfx^{(0)}\circ 
\bfx^{(1)}\circ \cdots\circ \bfx^{(k-1)}$, where $\abs*{\bfx^{(j)}} = 
Km$ for all~$j\in [k]$, and apply \cref{alg:torn-gray-encode} to 
$(\bfx^{(j)})_{j\in [k]}$ in succession; observe that every operation 
requires only $\ceilenv*{n/\lmin}$ distinct indices, and we utilize 
available indices in order throughout the $k$~operations.

We observe that the proofs of \cref{lem:suffix-iden,lem:torn-gray} 
hold without change, hence this amendment encodes into an $(\lmin,
\lmax)$-multistrand torn-paper code, which we denote 
$\code{cnst:torn-gray}(n,k)\in \cX_{n,k}$. Finally, following the 
proof of \cref{thm:torn-gray-red} we have 
\begin{IEEEeqnarray*}{+rCl+x*}
	\red(\code{cnst:torn-gray}(n,k)) 
	&=& k \big(K (\lmin - m(N)) + \lmin \>+ \\*
	&& \IEEEeqnarraymulticol{1}{r}{(n\bmod\lmin)\big)} \\
	&\leq&  \frac{n k}{a} \parenv*{1 + (1+o(1)) 
	\parenv*{\frac{f(n k)}{\log(n k)} + \frac{1}{f(n k)}}};
\end{IEEEeqnarray*}
As in \cref{thm:torn-gray-red}, for $f(n) = (1+o(1)) \sqrt{\log(n k)}$ 
we have 
\begin{IEEEeqnarray*}{+rCl+x*}
	\red(\code{cnst:torn-gray}(n,k)) &\leq& \frac{n k}{a} 
	\parenv*{1 + \frac{2+o(1)}{\sqrt{\log(n k)}}}.
\end{IEEEeqnarray*}
From \cref{lem:torn-lin-red} we have $\log\abs*{\cX_{n,k}} \geq 
(n-\log(k)) k$, concluding the proof.
\end{IEEEproof}
Again, by \cref{cor:torn-asym-rate,thm:torn-gray-red} the rate of the 
construction is asymptotically optimal.

\section{Error-Correcting Torn-paper Codes}\label{sec:torn-single-sub}

In this section, we extend the study of torn-paper codes to a noisy 
setup. We consider two models of noise. The first one assumes that 
the encoded string, before segmentation, suffers at most some $t$ 
substitution errors. The second model corresponds to the case where 
some of the segments are deleted during segmentation.

\subsection{Substitution-Correcting Torn-paper Codes}

For a string $\bfx$, its \emph{$t$-error torn-paper ball}, denoted 
by $\cB\cT_{\lmin}^{\lmax}(\bfx;t)$, is defined as the set of all 
possible $(\lmin,\lmax)$-segmentations after introducing at most 
$t$~errors to $\bfx$, that is,
\begin{align*}
    \cB\cT_{\lmin}^{\lmax}(\bfx;t) \deq 
    \bigcup_{\bfy\in B_t(\bfx)}\cT_{\lmin}^{\lmax}(\bfy),
\end{align*}
where $B_t(\bfx) = \mathset{\bfy}{d_H(\bfx,\bfy)\leq t}$ is the 
radius-$t$ Hamming ball centered at $\bfx$. A code $\cC$ is called a 
\emph{$t$-error single-strand torn-paper code} if for all $\bfx_1, 
\bfx_2\in \cC$, $\bfx_1\neq \bfx_2$, it holds that 
\begin{align*}
    \cB\cT_{\lmin}^{\lmax}(\bfx_1;t)\cap 
    \cB\cT_{\lmin}^{\lmax}(\bfx_2;t) = \emptyset.
\end{align*}

Our goal in this section is to show how to 
adjust~\cref{cnst:torn-gray} in order to produce $t$-error 
single-strand torn-paper codes. 
We first explain the main ideas of the required modifications. Let 
$\bfz = \enc[cnst:torn-gray](\bfx)\in \code{cnst:torn-gray}(n)$ 
(encoded with \cref{alg:torn-gray-encode}) and let $\cU\in 
\cB\cT_{\lmin}^{\lmax}(\bfz;t)$ be an $(\lmin,\lmax)$-segmentation of 
some word $\bfz'$, where $d_H(\bfz,\bfz')\leq t$. The main task of the 
noiseless decoder of $\code{cnst:torn-gray}(n)$ was to first calculate 
the index, and thus the location in $\bfz$, of every segment $\bfu\in 
\cU$. 
However, in the presence of errors, calculating the index of a segment 
$\bfu\in \cU$ based on the first (perhaps partial) occurrence of an 
encoded index within $\bfu$ might result with the misplacement of all 
the (perhaps partial) information blocks $\bfy_i$ that are contained 
in $\bfu$. Hence, a more careful approach is necessary for index 
decoding.

Before presenting our construction for $t$-error single-strand 
torn-paper codes, we introduce several additional required 
definitions. 
For a string $\bfu$, define $\cT_{\lmin}^+(\bfu)$ to be the multiset 
of non-overlapping $\lmin$-segments of $\bfu$, where the last segment 
is of length $\ell$, $\lmin\leq \ell < 2\lmin$. A segment $\bfw\in 
\cT_{\lmin}^+(\bfu)$ is called \emph{A-decodable} if, informally, 
\cref{alg:decode-index-lmin} returns (perhaps erroneous) output when 
given $\bfw$ as input. More formally, if $\bfw$ satisfies one of the 
following conditions.
\begin{enumerate}
\item 
$\bfw$ either contains a unique complete occurrence of $10^{f(n)}1$, 
or it doesn't contain complete occurrences but contains a cyclic 
occurrence (if $\lmin<\abs*{\bfw}<2\lmin$, require instead that either 
the $\lmin$-prefix or the $\lmin$-suffix of $\bfw$ contain a cyclic 
occurrence).
    
\item 
$\bfw$ contains precisely two complete occurrences of $10^{f(n)}1$, 
and there exist a unique pair of occurrences (either complete or 
complete-to-suffix/prefix) whose locations are at distance precisely 
$\lmin$.
Recall that $\bfw$ cannot contain more than two complete occurrences 
of $10^{f(n)}1$, except in the presence of errors, hence those cases 
can safely be discarded (see \cref{thm:torn-single-sub-1} in the proof 
of \cref{thm:torn-single-sub} for a formal proof).
\end{enumerate}

Let $\bfw$ be an A-decodable segment. Then, by definition, there is at 
least one occurrence (perhaps cyclic) of $10^{f(n)}1$ within $\bfw$ 
and, if there is more than a single occurrence, then there is exactly 
one pair of occurrences such that the difference between their 
locations is $\lmin$. Consider the $\alpha$-segments of $\bfw$ 
preceding these occurrences as encoded indices; if the (first) 
occurrence of $10^{f(n)}1$ in $\bfw$ is at location $\ell<\alpha$, 
concatenate the $(\alpha-\ell)$-segment of $\bfw$ at location $\lmin+
\ell-\alpha$, to the $\ell$-prefix of $\bfw$, and consider the 
resulting length-$\alpha$ string to be a \emph{cyclic encoded index}.

An A-decodable segment $\bfw$ is called \emph{valid} if, informally, 
there appears at least one `valid' encoded index in $\bfw$, and no 
conflicting pair of such indices. More formally, a valid segment 
$\bfw$ is an A-decodable segment that satisfies one of the following 
conditions:
\begin{enumerate}
\item 
$\bfw$ contains no complete encoded index, hence it contains only a 
cyclic encoded index.

\item 
$\bfw$ contains a single complete encoded index, and its parity 
symbol is correct.

\item 
$\bfw$ contains two complete encoded indices, and either exactly one 
of their parity symbols is correct, or both are correct and the 
indices are consecutive in the applied Gray code.
\end{enumerate}

\begin{construction}\label{cnst:torn-single-sub}
We construct a concatenated code, using~\cref{cnst:torn-gray} as 
inner-code, and an arbitrary $(K, q^{m(N) M}, 2t+1)_{q^{m(N)}}$ 
error-correcting code~$\code{EC}$, with an encoding algorithm 
$\enc_{\mathrm{EC}}\colon (\Sigma^{m(N)})^M \to (\Sigma^{m(N)})^K$, as 
outer-code (here, $K,N$ are the parameters of~\cref{cnst:torn-gray}). 
The resulting $t$-error $(\lmin,\lmax)$-single-strand torn-paper code 
is denoted $\code{cnst:torn-single-sub}(n)$, with the associated 
encoder $\enc[cnst:torn-single-sub]:\Sigma^{m(N) M}\to \Sigma^n$.
\end{construction}

We observe the following property of \cref{cnst:torn-single-sub}. 
Assume one retrieves a noisy version~$\bfz'$ of~$\bfz = 
\enc[cnst:torn-single-sub](\bfx)$, e.g., from any reconstruction 
algorithm; further assume that $\bfz,\bfz'$ agree on all locations 
containing encoded indices~$\bfc''_i$ or markers~$1 0^{f(n)} 1$ (as 
their locations in $\bfz$ are known a priori and do not depend on the 
information~$\bfx$). 
Thus, one extracts from $\bfz'$ (perhaps erroneous) encoded 
information blocks, denoted $\bfy'_i$. Denote by $e$ the number of 
encoded information blocks that were not recovered (e.g., due to 
conflicts in the reconstruction algorithm), and by $s$ the number of 
encoded blocks that were recovered incorrectly (i.e., $\bfy'_i\neq 
\bfy_i$). Since the information string $\parenv*{\bfx_i}_{i\in[M]}
\in\Sigma^{m(N) M}$ is encoded using a $(K, q^{m(N) M}, 
2t+1)_{q^{m(N)}}$ error-correcting code, it suffices that $2s+e\leq 2t$ 
to guarantee correct decoding.

In order to reconstruct a noisy version $\bfz'$ of $\bfz$, we define a 
modification of \cref{alg:decode-index-lmin}, as follows. First, given 
an $(\lmin,\lmax)$-segmentation $\cU\in\cB\cT_{\lmin}^{\lmax}(\bfz;t)$ 
we apply the reconstruction algorithm not directly to $\cU$, but 
rather to valid segments in 
\begin{align*}
	\cT^+_{\lmin}(\cU)\deq 
	\multset*{\cT^+_{\lmin}(\bfu)}{\bfu\in \cU}.
\end{align*}
Secondly, in case a valid $\bfw\in \cT^+_{\lmin}(\cU)$ contains 
multiple (perhaps cyclic) occurrences of an encoded index, the 
algorithm selects one to decode by prioritizing complete occurrences 
over cyclic ones, and in the case of complete occurrences, accepting 
the first containing a correct parity symbol (since $\bfw$ is valid, 
such occurrence exists in this case). Decoding of the selected encoded 
index is then performed as described in \cref{alg:decode-index-lmin}, 
and denoted by $\ind'(\bfw)$.

For an $(\lmin,\lmax)$-segmentation $\cU\in 
\cB\cT_{\lmin}^{\lmax}(\bfz;t)$, we define the set 
\begin{align*}
	\cZ(\cU)\deq \mathset*{(\ind'(\bfw),\bfw)}{\bfw\in 
	\cT^+_{\lmin}(\cU)\ \text{is valid}}.
\end{align*}
If $(j,\bfw), (j,\bfw')\in \cZ(\cU)$ for some~$j$ and $\bfw\neq\bfw'$, 
we define a restriction $\cZ'(\cU)$ of $\cZ(\cU)$ by including only 
the shortest, lexicographically-least, segment (i.e., $\cZ'(\cU)$ 
defines a proper function).

Given the set $\cZ'(\cU)$ we decode a string $\bfz'$ as follows.
\begin{enumerate}
\item 
Fill the encoded indices and the markers in $\bfz'$ in the correct 
locations as defined in~\cref{alg:torn-gray-encode} (note again that 
these locations do not depend on the information).

\item 
Next, we iterate over any pair $(\ind'(\bfw),\bfw)\in\cZ'(\cU)$ and 
update $\bfz'$ with the symbols of the encoded blocks $\bfy_i$'s 
within $\bfw$; If there is a collision of symbols in the same position 
within an encoded block $\bfy'_i$ for some~$i$, $i\in [K]$, we erase 
$\bfy'_i$ completely from $\bfz'$.

\item 
If an encoded block $\bfy'_i$ is partially filled at the end of the 
process (i.e., there are missing symbols within $\bfy'_i$) we erase 
the encoded block $\bfy'_i$. 
\end{enumerate}
The output~$\bfz'$ of this decoding procedure over the segmentation 
$\cU\in\cB\cT_{\lmin}^{\lmax}(\bfz;t)$ is denoted by 
$\dec[cnst:torn-single-sub](\cU)\deq \bfz'$.

We now prove that $\code{cnst:torn-single-sub}(n)$ is a $t$-error 
single-strand torn-paper code.

\begin{theorem}\label{thm:torn-single-sub}
Let $\bfz = \enc[cnst:torn-single-sub](\bfx)$, $\cU\in 
\cB\cT_{\lmin}^{\lmax}(\bfz;t)$, and let $\bfz' = 
\dec[cnst:torn-single-sub](\cU)$ be the noisy version of~$\bfz$ 
reconstructed by the aforementioned algorithm. 
Then, it holds that $2s+e\leq 2t$, where $e,s$ are defined as 
previously explained; i.e., any inner-channel error propagates as, at 
most, either one outer-channel error or two outer-channel erasures.
\end{theorem}
\begin{IEEEproof}
By definition, $\cU$ is obtained by first introducing up to $t$ errors 
to $\bfz$, and then performing an $(\lmin,\lmax)$-segmentation to the 
obtained word. For the rest of the proof, we fix an arbitrary $(\lmin,
\lmax)$-segmentation pattern, and for $\bfz\in \Sigma^n$ we denote 
$\cU\in \cT_{\lmin}^{\lmax}(\bfz)$ obtained from this pattern by $\cU 
= T(\bfz)$. In particular, observe for $\norm{\bfv}\leq t$ that 
$T(\bfz+\bfv)\in \cB\cT_{\lmin}^{\lmax}(\bfz;t)$.

For convenience, we denote by $\bfz'_\bfv\deq 
\dec[cnst:torn-single-sub](T(\bfz+\bfv))$, and by $e_\bfv$ 
(respectively, $s_\bfv$) the number of encoded information blocks 
$\bfy'_i$ in $\bfz'_\bfv$ which were not recovered (recovered 
erroneously). We shall prove the following proposition, which 
justifies the claim. Let $\bfz\deq \enc[cnst:torn-single-sub](\bfx)$, 
$\bfv\in \Sigma^n$ such that $\norm{\bfv}\leq t$. Then $e_\bfv + 
2s_\bfv\leq 2\norm{\bfv}$.

The proof is done by induction on $\norm{\bfv}$. First observe by 
\cref{lem:torn-gray} that $e_\bfzero = s_\bfzero = 0$ (here, $\bfzero$ 
is the all-zero string). 
For the induction step, assume that the claim holds for any 
$\bfv'\in \Sigma^n$, $\norm{\bfv'} < t$. Let $\bfv\in \Sigma^n$, 
$\norm{\bfv}=t$. 
Take any $\bfu'\in\cT^+_{\lmin}(T(\bfz+\bfv))$ 
affected by $t'>0$ errors. Decompose $\bfv = \bfv'+\bfv''$ such that 
$\norm{\bfv'}=t'$, $\norm{\bfv''}=t-t'$, and $\bfu'$ contains the 
support of $\bfv'$. 
Consider the decoder output $\bfz'_{\bfv''}$; by the induction 
assumption, 
\begin{align*}
    e_{\bfv''} + 2 s_{\bfv''} \leq 2 (t-t').
\end{align*}
We denote by $\bfu$ the segment corresponding to $\bfu'$ in 
$\bfz'_{\bfv''}$. Note that $\bfu$ contains no errors and its index is 
recovered correctly by the decoder. Hence, each encoded block that 
intersects $\bfu$ is either correct in $\bfz'_{\bfv''}$, or it is 
erased due to errors in other segments. 

Denoting by $\delta$ the number of encoded information blocks 
intersecting $\bfu$, we let 
\begin{enumerate*}[label=(\roman*)]
\item 
$\rho_1$ be the number of those recovered correctly 
in~$\bfz'_{\bfv''}$; 

\item 
$\rho_2$ be the number of those erased in~$\bfz'_{\bfv''}$ due to to 
collisions resulting from incorrect index-decoding in other segments; 
and 

\item 
$\rho_3$ be the number of those erased in~$\bfz'_{\bfv''}$ due to 
erasures of other, intersecting, segments.
\end{enumerate*}
Observe that $\delta = \rho_1+\rho_2+\rho_3\in \bracenv*{1,2,3}$, 
depending on $\abs*{\bfu}$ and its location).

The rest of the proof is done by cases.
\begin{enumerate}
\item \label{thm:torn-single-sub-1}
If $\bfu'$ is not valid, then all encoded information blocks 
intersecting $\bfu'$ are erased at the decoder. Hence, the $\rho_1$ 
correctly recovered blocks in $\bfz'_{\bfv''}$ which intersect $\bfu$ 
are erased in $\bfz'_\bfv$.

In addition, each of the $\rho_2$ blocks corresponding to blocks 
intersecting $\bfu$ which are erased due to collisions, might instead 
cause incorrect recovery of encoded information blocks in~$\bfu'$. 
Hence, $e_\bfv\leq e_{\bfv''}+\rho_1-\rho_2$ and $s_\bfv\leq 
s_{\bfv''} + \rho_2$, and we note 
\begin{align*}
	e_\bfv + 2 s_\bfv 
	&\leq (e_{\bfv''} + 2 s_{\bfv''}) + \rho_1 + \rho_2 \\
	&\leq 2 (t-t') + \delta = 2 \norm{u} - (2 t' - \delta).
\end{align*}
Since $t'\geq 1$, we have $e_\bfv + 2 s_\bfv\leq 2 \norm{u}$ unless 
$\delta = 3$; however, in that case $\bfu$ contains two complete 
instances of $1 0^{f(n)} 1$ whose locations are at distance $\lmin$, 
both preceded by complete occurrences of encoded indices, and since 
$\bfu'$ is not valid we have $t'>1$, which also concludes the proof. 

\item 
If $\bfu'$ is valid but its index is incorrectly decoded, then the 
$\rho_1$ encoded information blocks that are recovered correctly in 
$\bfz'_{\bfv''}$ are erased in $\bfz'_\bfv$, and $\rho_2$ encoded 
information blocks, corresponding to those intersecting $\bfu$ which 
are erased in $\bfz'_{\bfv''}$ due to collisions, might be recovered 
incorrectly in~$\bfz'_\bfv$.

Furthermore, the placement of $\bfu'$ at an incorrectly decoded 
location causes $\delta$ additional encoded information blocks to be 
either erased (due to collisions) or incorrectly recovered (where the 
correct blocks appear in invalid segments, i.e., are erased in 
$\bfz'_{\bfv''}$). Denoting the number of blocks of the former type by 
$\delta_1$, and the latter $\delta_2$, we then have $e_\bfv = 
e_{\bfv''} + \rho_1 - \rho_2 + \delta_1 - \delta_2$ and $s_\bfv\leq 
s_{\bfv''} + \rho_2 + \delta_2$. Hence, 
\begin{align*}
	e_\bfv + 2 s_\bfv 
	&\leq \parenv*{e_{\bfv''} + 2 s_{\bfv''}} 
	+ \rho_1 + \rho_2 + \delta_1 + \delta_2 \\
	&\leq 2 (t-t') + 2 \delta = 2 \norm{\bfv} - 2 (t' - \delta).
\end{align*}
To conclude, we require $t'\geq \delta$. Indeed, observe that if 
$\delta=2$ then $\bfu$ contains a complete occurrence of an encoded 
index followed by $1 0^{f(n)} 1$, requiring $t'\geq 2$ for incorrect 
recovery. 
Likewise, if $\delta=3$ then $\bfu$ contains two complete occurrences 
of encoded indices whose locations are at distance $\lmin$, each 
followed by $1 0^{f(n)} 1$; incorrect recovery of the index therefore 
requires at least two errors in one of them in addition to further 
errors in the other index or $1 0^{f(n)} 1$ marker, or an error in 
each $1 0^{f(n)} 1$ marker in addition to further errors to generate 
such a marker at an alternative location, hence $t'\geq 3$ as well.

\item 
Finally, if the index of $\bfu'$ is decoded correctly (and, in 
particular, $\bfu'$ is valid), then recalling that the index of $\bfu$ 
is also decoded correctly, we clearly have $e_\bfv = e_{\bfv''}$. 
Since any error in $\bfu'$ can cause an error in at most a single 
encoded information block, we have that $s_\bfv\leq s_{\bfv''}+t'$. 
Hence, 
\begin{IEEEeqnarray*}{+rCl+x*}
	e_\bfv + 2s_\bfv 
	&\leq& e_{\bfv''} + 2 (s_{\bfv''} + t') \\
	&=& \parenv*{e_{\bfv''} + 2 s_{\bfv''}} + 2 t'
	\leq 2 t = 2 \norm{\bfv}. &\IEEEQEDhere
\end{IEEEeqnarray*}
\end{enumerate}
\end{IEEEproof}

\begin{theorem}\label{thm:torn-sub-red}
Denote the redundancy of the outer-code $\cC_{EC}$ used in 
\cref{cnst:torn-single-sub} by $\rho_{EC}\deq K-M$. Then, operating 
$\enc[cnst:torn-gray]$ as in \cref{thm:torn-gray-red}, with $f(n) = 
(1+o(1)) \sqrt{\log(n)}$, we have 
\begin{IEEEeqnarray*}{+rCl+x*}
	\red(\code{cnst:torn-single-sub}(n)) 
    &\leq& \frac{n}{a} \bigg(1 + \frac{f(n)}{\log(n)} + \frac{1}{f(n)-1} \>+ \\*
    &&\hphantom{\frac{n}{a} \bigg(} \frac{9 + 2/(f(n)-1)}{\log(n)} + \frac{4a}{q^{f(n)}} + \frac{2 a^2 + 2}{n}\bigg) \>+ \\*
	&& \rho_{EC} \bigg((a-1) \log(n) - 2 \sqrt{\log(n)} \>- \\*
	&&\hphantom{\rho_{EC} \bigg(} 11 - \frac{3}{\sqrt{\log(n)}-1} 
	- \frac{4 a \log(n)}{q^{\sqrt{\log(n)}}}\bigg) \\
	&=& \frac{n}{a} \parenv*{1 + \frac{2+o(1)}{\sqrt{\log(n)}}} \>+ \\*
	&& \rho_{EC} \parenv*{(a-1) \log(n) - (2+o(1)) \sqrt{\log(n)}}.
\end{IEEEeqnarray*}
Furthermore, when $a>2$ then the outer-code $\cC_{EC}$ can be an MDS 
code and hence $\rho_{EC} = 2t$.
\end{theorem}
\begin{IEEEproof}
By \cref{cnst:torn-single-sub}, $\red(\code{cnst:torn-single-sub}(n)) 
= \red(\code{cnst:torn-gray}(n))+\rho_{EC} \cdot m(N)$.

We recall from the proof of \cref{thm:torn-gray-red} that for $f(n) = 
\ceilenv[\big]{\sqrt{\log(n)}}$ it holds that 
\begin{IEEEeqnarray*}{+rCl+x*}
	m(N) &\geq& \lmin - \log(n) - f(n) - \frac{\log(n)}{f(n)-1} \>- \\*
	&& \IEEEeqnarraymulticol{1}{r}{9 - \frac{2}{f(n)-1} - 
	\frac{4 a \log(n)}{q^{f(n)}}} \\
	&\geq& (a-1) \log(n) - 2 \sqrt{\log(n)} \>- \\*
	&& \IEEEeqnarraymulticol{1}{r}{11 - \frac{3}{\sqrt{\log(n)}-1} - 
	\frac{4 a \log(n)}{q^{\sqrt{\log(n)}}}}, 
\end{IEEEeqnarray*}
satisfying the former part of claim.

Next, for $a>2$ we observe that $m(N) > \log(n) - \log\log(n) + O_n(1) 
= \log(K)$, implying that an RS code may be used in 
\cref{cnst:torn-single-sub}, satisfying the latter part. 
\end{IEEEproof}

Before concluding the section, we outline an extension of 
\cref{cnst:torn-single-sub} to the case $k>1$, i.e., to 
\emph{$t$-error multi-strand torn-paper codes}.

\begin{corollary}
Take $n,k$ such that $k>1$, $\log(k)=o(n)$; let $\lmin = 
\ceilenv*{a \log(n k)}$, for $a>1$, and take some $\lmax\geq \lmin$. 
Amend \cref{cnst:torn-single-sub} as was done in 
\cref{thm:multistrand} to \cref{cnst:torn-gray}, using a $(k K, 
q^{m(N) M}, 2t+1)_{q^{m(N)}}$ error-correcting code~$\code{EC}$, with 
redundancy $\rho_{EC}\deq k K - M$. 
Then the resulting code~$\code{cnst:torn-single-sub}(n,k)$ is a 
$t$-error $(\lmin, \lmax)$-multistrand torn-paper code, satisfying 
\begin{IEEEeqnarray*}{+rCl+x*}
	\red\parenv*{\code{cnst:torn-single-sub}(n,k)} 
	&\leq& \frac{n k}{a} \parenv*{1+\frac{2+o(1)}{\sqrt{\log(n k)}}} \>+ \\*
	&& \rho_{EC} \Big((a-1) \log(n k) \>- \\*
	&& \phantom{\rho_{EC} \Big(} (2+o(1)) \sqrt{\log(n k)}\Big).
\end{IEEEeqnarray*}
\end{corollary}
\begin{IEEEproof}
The proof of \cref{thm:torn-single-sub} applies without change. 
As in \cref{thm:multistrand}, we have 
\begin{align*}
	m(N) &= (a-1) \log(n k) - (1+o(1)) \parenv*{f(n k) + 
	\frac{\log(n k)}{f(n k)}},
\end{align*}
and following the steps of \cref{thm:torn-sub-red}, we have the 
claimed upper bound on redundancy, for $f(n) = (1+o(1)) 
\sqrt{\log(n k)}$.
\end{IEEEproof}

\subsection{Deletion-Correcting Torn-paper Codes}

For a string $\bfx$, its \emph{$t$-deletion torn-paper ball}, 
$\cD\cT_{\lmin}^{\lmax}(\bfx;t)$, is defined as all the subsets with 
at most $t$ missing segments of all the possible 
$(\lmin,\lmax)$-segmentations of $\bfx$, that is,
\begin{align*}
	\cD\cT_{\lmin}^{\lmax}(\bfx;t) 
	\deq \bigcup_{\mathclap{S\in \cT_{\lmin}^{\lmax}(\bfx)}} 
	\mathset*{S' \subseteq S}{\abs*{S}-\abs*{S'}\leq t}.
\end{align*}
A code $\cC$ is called a \emph{$t$-deletion torn-paper code} if for 
all $\bfx_1,\bfx_2\in \cC$ it holds that 
$\cD\cT_{\lmin}^{\lmax}(\bfx_1;t) \cap 
\cD\cT_{\lmin}^{\lmax}(\bfx_2;t) = \emptyset.$

In this section, we utilize \emph{burst-erasure-correcting (BEC) 
codes} in our constructions, which are defined next. For a string 
$\bfx$, its \emph{$t$-burst $L$-erasures ball}, denoted by 
$\ball{BE}^L(\bfx;t)$, is defined as the set of all strings that can 
be obtained from $\bfx$ by at most $t$ burst of erasures, each of 
length at most $L$. A code $\cC$ is called a \emph{$t$-burst 
$L$-erasure correcting code} if for all $\bfx_1, \bfx_2\in \cC$, 
$\ball{BE}^L(\bfx_1;t)\cap \ball{BE}^L(\bfx_2;t) = \emptyset$.

Next, we present a generic construction of $t$-deletion torn-paper 
codes. Let $\whlmax\deq \lmax - \ceilenv[\big]{\frac{\lmax}{\lmin}} 
(\alpha + f(n) + 2)$. This construction is based on 
\cref{cnst:torn-gray} and assumes the existence of a systematic linear 
$t$-burst $\whlmax$-erasure correcting code, denoted by $\code{BEC}$.

\begin{construction}\label{cnst:torn-single-tdelnew}
Let $\rho > 0$ be an integer that is determined next. This 
construction uses the following family of codes:

\emph{Systematic BEC encoding}. Let $\enc_{\rm BEC}\colon 
\Sigma^{(K-\rho)N}\to \Sigma^{\rho_{\rm BEC}}$ denote the 
systematic encoder of the code $\code{BEC}$, such that for any string 
$\bfv\in \Sigma^{(K-\rho) N}$, $\bfv \circ \enc_{\rm BEC}(\bfv) 
\in \code{BEC}$ (for convenience we assume that $\enc_{\rm BEC}(\bfv)$ 
returns only the encoded systematic redundancy symbols). The 
redundancy of this encoder is denoted by $\rho_{\rm BEC}$. The 
parameter $\rho$ is defined $\rho\deq \Big\lceil \frac{1}{N} 
\Big\lfloor \rho_{\rm BEC}\cdot \frac{f(n)}{f(n)-1} 
\Big\rfloor \Big\rceil$.

Next, we utilize a generalized concatenated coding approach, where 
\cref{cnst:torn-gray} is used as inner-code for $K-\rho$ information 
blocks, and with a slight adjustment also for the $\rho$ redundant 
blocks, as follows: 
\begin{enumerate}
\item \emph{The length of the input string $\bfx$}. 
The input of this construction is $\bfx\in \Sigma^{(K-\rho) m(N)}$. 
That is, this construction has additional redundancy of $\rho m(N)$ 
symbols compared to \cref{cnst:torn-gray}. The input string is divided 
to $K-\rho$ information blocks each of length~$m(N)$, denoted by 
$\bfx_0, \ldots, \bfx_{K-\rho-1}$.

\item \emph{The generation of the encoded blocks $\bfy_i$'s}. 
The first $K-\rho$ blocks are generated from the corresponding 
$\bfx_i$'s using the RLL encoder $E_m^{\rm RLL}$ similarly to 
\cref{cnst:torn-gray}. Let $\bfy^*\deq \bfy_0 \circ \cdots \circ 
\bfy_{K-\rho-1} \in \Sigma^{(K-\rho) N}$ denote their 
concatenation. Next, we apply  $\enc_{\mathrm{BEC}}$ to obtain $\bfw 
\deq \enc_{\rm BEC}(\bfy^*)$, and denote by $\bfw^*$ the result of 
inserting `$1$'s into $\bfw$ at every location divisible by $f(n)$ (in 
particular, $\bfy^*\circ \bfw^*$ does not contain a length-$f(n)$ 
zero-run). 
Then, $\bfw^*$ is divided to the remaining segments $\bfy_{K-\rho}, 
\ldots, \bfy_{K-1} \in \Sigma^N$ (if $\abs*{\bfw^*}$ is not a 
multiple of~$N$, $\bfy_{K-1}$ is padded with $1$'s to length~$N$). 
Note that the parameter $\rho$ satisfies $\rho N\geq 
\Big\lfloor \rho_{\rm BEC}\cdot \frac{f(n)}{f(n)-1}\Big\rfloor = 
\abs*{\bfw^*}$, 
hence one may continue to follow the steps of \cref{cnst:torn-gray} 
without change.
\end{enumerate}
We now indeed continue identically to \cref{cnst:torn-gray}. That is, 
an index and a marker are appended to the beginning of each encoded 
block~$\bfy_i$ to construct a segment~$\bfz_i$ of length~$\lmin$. 
Then, $\bfz_0,\dots,\bfz_{K-1}$ are concatenated along with 
$\bfz_K \circ  0^{n\bmod\lmin} = \bfc''_K \circ 1 0^{f(n)} 1 0^{N 
+ (n\bmod\lmin)}$ to obtain the encoded output string~$\bfz\in 
\Sigma^n$.
\end{construction}

Let $\code{del}(n)$ denote the constructed code. The correctness of 
\cref{cnst:torn-single-tdelnew} and redundancy calculation are proved 
in the next theorem.

\begin{theorem}\label{thm:deletions}
$\code{del}(n)$ is a $t$-deletion torn-paper code. Furthermore, it 
holds that 
\begin{IEEEeqnarray*}{+rCl+x*}
	\red(\code{del}(n)) &=& \red(\code{cnst:torn-gray}(n)) \>+ \\*
	&& m(N) \bigg\lceil \frac{1}{N} \bigg\lfloor \rho_{\rm BEC}\cdot 
	 \frac{f(n)}{f(n)-1} \bigg\rfloor \bigg\rceil.
\end{IEEEeqnarray*}
\end{theorem}
\begin{IEEEproof}
Let $\bfz\in \code{del}(n)$ be the encoded codeword of the input 
string $\bfx$, and take $\cU\in \cD\cT_{\lmin}^{\lmax}(\bfz;t)$. We 
shall prove that one can uniquely decode $\bfx$.

From \cref{lem:torn-gray}, for every $\bfu\in \cU$ which is not a 
substring of the suffix of length $(n \mod \lmin) + N+f(n)$ of 
$\bfz$, its index $\ind(\bfu)$ can be decoded using 
\cref{alg:decode-index-lmin}. The string~$\bfz$ can then be 
reconstructed by the locations of each received segment, with some 
segments erased (at identifiable locations). Let $\bfz'\in (\Sigma 
\cup \bracenv*{?})^n$ denote this partially reconstructed string, 
where `$?$' stands for erased symbols.

From the definition of $\cD\cT_{\lmin}^{\lmax}(\bfz;t)$, at most $t$ 
segments of $\bfz$ are missing from $\cU$. Therefore, $\bfz'\in 
\ball{BE}^{\lmax}(\bfz;t)$. By removing coordinates of $\bfz'$ 
corresponding to indices or markers (including '?' symbols), a string 
$\bfy'\in \ball{BE}^{\whlmax}\parenv*{\bfy_0\circ \cdots\circ 
\bfy_{K-1};t}$ is obtained, since there are at most $ \lmax - 
\whlmax = \ceilenv[\big]{\frac{\lmax}{\lmin}} (\alpha + f(n) + 2)$ 
symbols in any $\lmax$-segment of $\bfz$ belonging to either index 
or marker.

Finally, we remove from $\bfy'$ coordinates corresponding to `$1$'s 
inserted into $\bfw^*$; thus, we obtain a string $\widehat{\bfy}\in 
\ball{BE}^{\whlmax}(\bfy^* \circ \bfw; t)$. 
A decoder for $\code{BEC}$ may be invoked on $\widehat{\bfy}$ to 
retrieve $\bfy^* = \bfy_0\circ \cdots\circ \bfy_{K-\rho-1}$, and 
consequently $\bfx$ is obtained by applying the RLL decoder to each 
$\bfy_i$, $i\in [K-\rho]$.

To conclude the proof we observe that the asserted redundancy follows 
by definition, as precisely $\rho m(N)$ less information symbols are 
input at the encoder, in comparison to \cref{cnst:torn-gray}.
\end{IEEEproof}

Next, we note that an extension to the case $k>1$, i.e., to 
\emph{$t$-deletion multi-strand torn-paper codes}, is again 
straightforward.

\begin{corollary}
Amending \cref{cnst:torn-single-tdelnew}, one constructs a 
$t$-deletion multi-strand torn-paper code~$\code{del}(n, k)$ with 
redundancy 
\begin{IEEEeqnarray*}{+rCl+x*}
	\red(\code{del}(n,k)) &=& \red(\code{cnst:torn-gray}(n,k)) \>+ \\*
	&& m(N) \bigg\lceil \frac{1}{N} \bigg\lfloor \rho_{\rm BEC}\cdot 
	 \frac{f(n)}{f(n)-1} \bigg\rfloor \bigg\rceil.
\end{IEEEeqnarray*}
\end{corollary}
\begin{IEEEproof}
Here, an information string $\bfx\in \Sigma^{(kK - \rho) m(N)}$ is 
encoded with $E_m^{\rm RLL}$ into $\bfy^*$, and $\bfw^*$ is obtained 
utilizing a systematic BEC encoder on strings in $\Sigma^{(k K - \rho) 
N}$. It is segmented into $\bfy_{k K - \rho}, \ldots, \bfy_{k K}
\in\Sigma^{N}$; again, observing $\rho N\geq \rho_{\rm BEC}$ 
assures that this is possible. 
Then, each $K$~segments $\bfy_j$ are encoded, in order, with the 
remaining steps of \cref{alg:torn-gray-encode}, where again $I\deq 
\ceilenv*{\log\parenv*{k \ceilenv*{n/\lmin}}}$ and $\lmin = 
\ceilenv*{a \log(n k)}$, and indices are utilized by each operation in 
succession. It is straightforward that the proof of 
\cref{thm:deletions} can be followed to show that $\code{del}(n,k)$ is 
a $t$-deletion multi-strand torn-paper code, with the above 
redundancy.
\end{IEEEproof}

Before concluding the section, we discuss the cases of $t\in 
\bracenv*{1,2}$, in which more is known on the construction of BEC 
codes.

For $t=1$, we use a systematic interleaving parity BEC code as the 
code $\code{BEC}$. Namely, the redundancy string $\bfw = 
\enc_{\rm BEC}(\bfy^*)$ is of length $\rho_{\rm BEC}=\whlmax$, and 
\begin{align*}
	w_i\deq \sum\Big._{k\in \sparenv*{\ceilenv*{\frac{(K-\rho) N 
	- i}{\whlmax}}}} y^*_{i + k \whlmax}
\end{align*}
for all $i\in [\whlmax]$, i.e., $w_i$ is a single parity symbol for 
$\parenv*{y^*_i, y^*_{i+\whlmax}, \ldots}$. Denote this code by 
$\code{del,1}$.

For $t = 2$, we state for completeness the following basic proposition 
which draws the connection between burst-error-correcting codes and 
burst-erasure-correcting codes. We note that this fact has been 
mentioned before in~\cite{ChiBahTan69}, for a single burst of errors.

\begin{lemma}\label{eq:burst}
For $0<\ell\leq n$ and $\bfx,\bfy\in \Sigma^n$, it holds that $\bfx, 
\bfy$ are confusable under $t$~bursts of errors of lengths at 
most~$\ell$ if and only if they are confusable under $2t$ bursts of 
erasures of lengths at most~$\ell$.
\end{lemma}
\begin{IEEEproof}
Denote $\bfx = (x_j)_{j\in [n]}$, $\bfy = (y_j)_{j\in [n]}$, and 
$I_i\deq \bigcup_{j\in[t]} (k^{(i)}_j + [\ell])$, for $i=0,1$ and some 
$\bracenv[\big]{k^{(i)}_j}_{j\in[t]}\subseteq[n]$. Assume there exist 
$\bfe^{(0)},\bfe^{(1)}\in \Sigma^n$ such that $\bfx + \bfe^{(0)} = 
\bfy + \bfe^{(1)}$, and $\supp(\bfe^{(i)})\subseteq I_i$, $i = 1,2$.
Then, one observes that $\bfx_{[n]\setminus(I_0\cup I_1)} = 
\bfy_{[n]\setminus(I_0\cup I_1)}$.

Conversely, assume $\bfx_{[n]\setminus I} = \bfy_{[n]\setminus I}$, 
where $I\subseteq \bigcup_{j\in[2t]} (k_j + [\ell])$ for some 
$\bracenv[\big]{k_j}_{j\in[2t]}\subseteq[n]$, and without loss of 
generality $\bracenv[\big]{k_j}_{j\in[2t]}$ are increasing, and 
$k_j\leq k_{t+1} - \ell$ for all $j\leq t$. 
Let $I_i\deq \bigcup_{j\in\parenv*{it+[t]}} (k_j + [\ell])$ for $i = 
1,2$, and observe that $I_0\cup I_1=I$, $I_0\cap I_1=\emptyset$. For 
$i= 1,2$ and $j\in [n]$, let 
\begin{align*}
	e^{(i)}_j\deq \begin{cases}
	(-1)^i \parenv*{y_j-x_j}, & j\in I_i; \\
	0, & \text{otherwise}.
	\end{cases}
\end{align*}
Then, denoting $\bfe^{(i)} = (e^{(i)}_j)_{j\in [n]}$ for $i = 1,2$, we 
have $\bfx + \bfe^{(0)} = \bfy + \bfe^{(1)}$, which completes the 
proof.
\end{IEEEproof}

A construction of $2$-deletion torn-paper codes is derived from 
\cref{cnst:torn-single-tdelnew}, using a BEC code for $t=2$. Hence, 
by \cref{eq:burst} one may use an $\whlmax$-burst error-correcting 
code. Observe that \cref{cnst:torn-single-tdelnew} requires a 
systematic encoder, which is guaranteed by several prior works with 
redundancy at most $\log((K-\rho) N) + \whlmax$; see, 
e.g.~\cite{AbdMcEOdlvTi86, Abd88}. These constructions require the 
alphabet~$\Sigma$ to be a field, and are linear and cyclic, which 
ensures the existence of a systematic encoder. For simplicity of 
derivation we bound this redundancy (from above) by $\whlmax+\log(n)$. 
Let $\code{del,2}$ denote this code.

The next corollary summarizes these results. For convenience, denote 
the difference 
\begin{align*}
	\Delta\red(\cC(n))\deq 
	\red(\cC(n)) - \red(\code{cnst:torn-gray}(n)),
\end{align*}
for a $t$-deletion torn-paper code~$\cC(n)\subseteq \Sigma^n$.

\begin{corollary}
For a prime power~$q$ and all admissible values of $n$ and $f(n)$ in 
\cref{cnst:torn-gray}, where $f(n)=\omega(1)$, $f(n)=o(\log(n))$ and 
with the RLL encoders of \cite{LevYaa19,YehBarMarYaa23}, it holds that 
\begin{align*}
	\Delta\red(\code{del,1}(n)) 
	&\leq \whlmax \cdot \frac{f(n)}{f(n)-1}, \\
	\Delta\red(\code{del,2}(n)) 
	&\leq (\whlmax+\log(n)) \cdot \frac{f(n)}{f(n)-1}.
\end{align*}
In particular, for $f(n) = (1+o(1)) \sqrt{\log(n)}$, 
\begin{align*}
	\Delta\red(\code{del,1}(n)) 
	&\leq  \whlmax\parenv*{1+\frac{1-o(1)}{\sqrt{\log(n)}}}, \\
	\Delta\red(\code{del,2}(n)) 
	&\leq (\whlmax + \log(n))\parenv*{1 + 
	\frac{1-o(1)}{\sqrt{\log(n)}}} .
\end{align*}
\end{corollary}

Note that if $\lmax = o(n)$ the asymptotic rate of $\code{del,1}(n)$ 
and $\code{del,2}(n)$ is asymptotically equal to the rate of 
$\code{cnst:torn-gray}(n)$. Thus, efficient encoding and decoding of 
$t$-deletion torn-paper codes, $t = 1,2$, is possible at rates 
arbitrarily close to the optimum.

\section{Conclusion}\label{sec:conc}

In this paper, we study the adversarial torn-paper channel, for which 
we present fundamental bounds and code constructions. We further study 
several extensions of this model, including multi-strand storage, 
substitution errors, or incomplete coverage. Importantly, our proposed 
constructions have linear-run-time encoders and decoders, and the 
resulting codes achieve asymptotically optimal rates.

We mention again that the adversarial model we assume in this work is 
chosen to enable analysis in the worst-case. More realistically, an 
adversarial channel where the \emph{average} of the received segments' 
lengths is bounded from below might be analyzed; unfortunately, this 
channel turns out to be hard to analyze in the worst-case, and such 
analysis is left for future works. It will be remarked that by the 
same methods of~\cref{lem:torn-lin-red}, it can be shown that the 
capacity of such an adversarial channel is bounded from above by 
$1-\frac{1}{a}$, where the lower bound on the average segment length 
is chosen to be $a \log(n)$. 
Coding for this channel appears to be more challenging; we point to 
the fact that an adversary is able to segment a non-vanishing fraction 
of the channel input into short substrings as a likely reason for that 
difficulty.

A naive solution, where the lower bound on the average segment length 
is $a \log(n)$ and $a>\frac{q}{q-1}$, is to apply 
\cref{cnst:torn-single-tdelnew} with parameter $a'$ satisfying $1 < a' 
< (1-\frac{1}{q}) a$; the decoder then discards any received segment 
shorter than $a' \log(n)$, creating at most $\frac{n}{a \log(n)}$ 
bursts of erasures of lengths at most $(a'-1) \log(n)$ (in the 
reconstructed information sequence). To recover the information, a BEC 
code~$\code{BEC}$ is used; since $\frac{1}{K m(N)} (a'-1) \log(n) 
\frac{n}{a \log(n)} = \frac{(a'-1)/a}{K m(N) / n} = \frac{a'}{a}+o(1) < 
1-\frac{1}{q}$, a positive-rate BEC code exists for all $a'$ in the 
permissible range (since positive-rate $\frac{a'-1}{a} 
n$-erasure-correcting codes exist in $\Sigma^{K m(N)}$); hence $a'$ can 
be optimized according to \cref{thm:deletions}, i.e., to maximize the 
achieved rate of $(1+o_n(1)) \parenv*{1-\frac{1}{a'}}\cdot 
R(\code{BEC})$. (Naively, one might utilize codes correcting 
$\frac{n}{a \log(n)}\cdot (a'-1) \log(n)$ erasures; by the GV bound, 
the achievable rate of this construction is at least $(1+o_n(1)) 
\parenv*{1-\frac{1}{a'}} \parenv[\big]{1-H_q(\frac{a'}{a})}$.)
Alternatively, if any integer $a'$ falls within the given range, 
\cref{cnst:pilot} can also be used with parameter~$a'$, where again 
segments shorter than $(a'+o(1)) \log(n)$ are discarded at the 
decoder, and reconstructed based on a BEC code~$\code{BEC}$ correcting 
up to $\frac{n}{a \log(n)}$ bursts of erasures of lengths at most 
$(1+o(1)) \log(n)$ in $\Sigma^{n/a'}$; however, the achieved rate of 
this construction is similarly $\parenv*{1-\frac{1}{a'}}\cdot 
R(\code{BEC})$, and applicable BEC codes are equivalent (i.e., they 
correct the same number of bursts, of length $(1+o(1)) \log(n)$ in 
$\Sigma^{n/a'}$ instead of length $(a'-1) \log(n)$ in $\Sigma^{K m(N)} 
= \Sigma^{(1-1/a'+o(1)) n}$).

Finally, for future research, we believe that applying our methods to 
a generalized channel, including multiple sources of noise 
concurrently, one may achieve similar results. Studying the channel 
under edit-errors, including insertions/deletions in addition to 
substitutions, is also of great interest for applications to DNA data 
storage.

\section*{Acknowledgments}

The authors gratefully acknowledge the valuable insight and advice 
offered to us by the two anonymous reviewers and associate editor, 
which were instrumental in streamlining the presentation of this paper.

\begin{IEEEbiographynophoto}{Daniella Bar-Lev}
(S'20) received the B.Sc. degree in computer science and the B.Sc. 
degree in mathematics in~2019, and the M.Sc. degree in computer science 
in~2021, from the Technion---Israel Institute of Technology, Haifa, 
Israel, where she is currently pursuing the Ph.D. degree with the 
Computer Science Department. 
Her research interests include algorithms, discrete mathematics, coding 
theory, and DNA storage.
\end{IEEEbiographynophoto}

\begin{IEEEbiographynophoto}{Sagi Marcovich}
(S'20) received the B.Sc. degree in software engineering and the M.Sc. 
degree in computer science from the Technion---Israel Institute of 
Technology, Haifa, Israel, in~2016 and~2021, respectively, where he is 
currently pursuing the Ph.D. degree with the Computer Science 
Department. 
His research interests include algorithms, information theory, and 
coding theory with applications to DNA based storage.
\end{IEEEbiographynophoto}

\begin{IEEEbiographynophoto}{Eitan Yaakobi}
(S'07--M'12--SM'17) received the B.A. degree in computer science and the 
B.A degree in mathematics in~2005, and the M.Sc. degree in computer 
science in~2007, from the Technion---Israel Institute of Technology, 
Haifa, Israel, and the Ph.D. degree in electrical engineering in~2011, 
from the University of California at San Diego, San Diego, CA, USA. 
From~2011 to~2013, he was a Post-Doctoral Researcher with the Department 
of Electrical Engineering, California Institute of Technology, and the 
Center for Memory and Recording Research, University of California at 
San Diego. Since~2016, he has been with the Center for Memory and 
Recording Research, University of California at San Diego. Since~2018, 
he has been with the Institute of Advanced Studies, Technical University 
of Munich, where he holds a four-year Hans Fischer Fellowship, funded by 
the German Excellence Initiative and the EU 7th Framework Program. He is 
currently an Associate Professor with the Computer Science Department, 
Technion---Israel Institute of Technology. He also holds a courtesy 
appointment with the Electrical and Computer Engineering (ECE) 
Department, Technion---Israel Institute of Technology. 
His research interests include information and coding theory with 
applications to non-volatile memories, associative memories, DNA 
storage, data storage and retrieval, and private information retrieval. 
He was a recipient of several grants, including the ERC Consolidator 
Grant. He received the Marconi Society Young Scholar in~2009 and the 
Intel Ph.D. Fellowship during~2010-2011. 
Since~2020, he has been serving as an Associate Editor for Coding 
and Decoding for the \textsc{IEEE Transactions on Information Theory}. 
\end{IEEEbiographynophoto}

\begin{IEEEbiographynophoto}{Yonatan Yehezkeally}
(S'12--M'20)
received the B.Sc.~degree (\emph{cum laude}) in mathematics and the 
M.Sc.~(\emph{summa cum laude}) and Ph.D. degrees in electrical and 
computer engineering from Ben-Gurion University of the Negev, 
Beer-Sheva, Israel, in~2013, 2017, and~2020 respectively. 
He is currently a Carl Friedrich von Siemens Post-Doctoral Research 
Fellow of the Alexander von Humboldt Foundation, with the 
Associate Professorship of Coding and Cryptography (Prof. 
Wachter-Zeh), School of Computation, Information and Technology, 
Technical University of Munich. 
His research interests include coding for novel storage media, with a 
focus on DNA-based storage and nascent sequencing technologies, as 
well as combinatorial structures and finite group theory.
\end{IEEEbiographynophoto}


\begin{thebibliography}{10}
\providecommand{\url}[1]{#1}
\csname url@samestyle\endcsname
\providecommand{\newblock}{\relax}
\providecommand{\bibinfo}[2]{#2}
\providecommand{\BIBentrySTDinterwordspacing}{\spaceskip=0pt\relax}
\providecommand{\BIBentryALTinterwordstretchfactor}{4}
\providecommand{\BIBentryALTinterwordspacing}{\spaceskip=\fontdimen2\font plus
\BIBentryALTinterwordstretchfactor\fontdimen3\font minus
  \fontdimen4\font\relax}
\providecommand{\BIBforeignlanguage}[2]{{%
\expandafter\ifx\csname l@#1\endcsname\relax
\typeout{** WARNING: IEEEtranS.bst: No hyphenation pattern has been}%
\typeout{** loaded for the language `#1'. Using the pattern for}%
\typeout{** the default language instead.}%
\else
\language=\csname l@#1\endcsname
\fi
#2}}
\providecommand{\BIBdecl}{\relax}
\BIBdecl

\bibitem{Abd88}
K.~A.~S. {Abdel-Ghaffar}, ``On the existence of optimum cyclic burst correcting
  codes over {GF}(q),'' \emph{IEEE Trans.~on Inform.~Theory}, vol.~34, no.~2,
  pp. 329--332, Mar. 1988.

\bibitem{AbdMcEOdlvTi86}
K.~A.~S. {Abdel-Ghaffar}, R.~J. {McEliece}, A.~M. {Odlyzko}, and H.~C.~A. {van
  Tilborg}, ``On the existence of optimum cyclic burst-correcting codes,''
  \emph{IEEE Trans.~on Inform.~Theory}, vol.~32, no.~6, pp. 768--775, Nov.
  1986.

\bibitem{AchDasMilOrlPan15}
J.~{Acharya}, H.~{Das}, O.~{Milenkovic}, A.~{Orlitsky}, and S.~{Pan}, ``String
  reconstruction from substring compositions,'' \emph{SIAM J.~Discrete Math.},
  vol.~29, no.~3, pp. 1340--1371, 2015.

\bibitem{Bal13}
F.~Balado, ``Capacity of {DNA} data embedding under substitution mutations,''
  \emph{IEEE Trans.~on Inform.~Theory}, vol.~59, no.~2, pp. 928--941, Feb.
  2013.

\bibitem{BorLopCarCezSeeStr16}
J.~{Bornholt}, R.~{Lopez}, D.~M. {Carmean}, L.~{Ceze}, G.~{Seelig}, and
  K.~{Strauss}, ``A {DNA}-based archival storage system,'' \emph{ACM SIGPLAN
  Notices}, vol.~51, no.~4, pp. 637--649, Mar. 2016.

\bibitem{BreBreTse13}
G.~{Bresler}, M.~{Bresler}, and D.~{Tse}, ``Optimal assembly for high
  throughput shotgun sequencing,'' \emph{{BMC} Bioinformatics}, vol.~14, no.~5,
  p. S18, Jul. 2013.

\bibitem{ChiBahTan69}
R.~T. {Chien}, L.~R. {Bahl}, and D.~{Tang}, ``Correction of two erasure bursts
  (corresp.),'' \emph{IEEE Trans.~on Inform.~Theory}, vol.~15, no.~1, pp.
  186--187, Jan. 1969.

\bibitem{Chinetal13}
C.-S. {Chin}, D.~H. {Alexander}, P.~{Marks}, A.~A. {Klammer}, J.~{Drake},
  C.~{Heiner}, A.~{Clum}, A.~{Copeland}, J.~{Huddleston}, E.~E. {Eichler},
  S.~W. {Turner}, and J.~{Korlach}, ``Nonhybrid, finished microbial genome
  assemblies from long-read {SMRT} sequencing data,'' \emph{Nature Methods},
  vol.~10, no.~6, pp. 563--569, Jun. 2013.

\bibitem{ChuGaoKos12}
G.~M. Church, Y.~Gao, and S.~Kosuri, ``Next-generation digital information
  storage in {DNA},'' \emph{Science}, vol. 337, no. 6102, pp. 1628--1628, 2012.

\bibitem{dnaalliancewp21}
\BIBentryALTinterwordspacing
{Contributing Members}, ``Preserving our digital legacy: an introduction to
  {DNA} data storage,'' The {DNA} Data Storage Alliance, White Paper, Jun.
  2021. [Online]. Available:
  \url{https://dnastoragealliance.org/dev/wp-content/uploads/2021/06/DNA-Data-Storage-Alliance-An-Introduction-to-DNA-Data-Storage.pdf}
\BIBentrySTDinterwordspacing

\bibitem{deBvAE51}
N.~G. {de Bruijn} and T.~{van Aardenne-Ehrenfest}, ``Circuits and trees in
  oriented linear graphs,'' \emph{Simon Stevin}, vol.~28, pp. 203--217, 1951.

\bibitem{EliGabMedYaa21}
O.~{Elishco}, R.~{Gabrys}, M.~{M\'{e}dard}, and E.~{Yaakobi}, ``Repeat-free
  codes,'' \emph{IEEE Trans.~on Inform.~Theory}, vol.~67, no.~9, pp.
  5749--5764, Sep. 2021.

\bibitem{ErlZie17}
Y.~{Erlich} and D.~{Zielinski}, ``{DNA} fountain enables a robust and efficient
  storage architecture,'' \emph{Science}, vol. 355, no. 6328, pp. 950--954,
  Mar. 2017.

\bibitem{GabMil19}
R.~{Gabrys} and O.~{Milenkovic}, ``Unique reconstruction of coded strings from
  multiset substring spectra,'' \emph{IEEE Trans.~on Inform.~Theory}, vol.~65,
  no.~12, pp. 7682--7696, Dec. 2019.

\bibitem{GanMosRac16}
S.~{Ganguly}, E.~{Mossel}, and M.~{Racz}, ``Sequence assembly from corrupted
  shotgun reads,'' in \emph{Proceedings of the 2016 {IEEE} International
  Symposium on Information Theory ({ISIT}), Barcelona, Spain}, Jul. 2016, pp.
  265--269.

\bibitem{GolCheDesLePSipBir13}
N.~{Goldman}, P.~{Bertone}, S.~{Chen}, C.~{Dessimoz}, E.~M. {LeProust},
  B.~{Sipos}, and E.~{Birney}, ``Towards practical, high-capacity,
  low-maintenance information storage in synthesized {DNA},'' \emph{Nature},
  vol. 494, no. 7435, pp. 77--80, Feb. 2013.

\bibitem{GraHecPudPauSta15}
R.~N. {Grass}, R.~{Heckel}, M.~{Puddu}, D.~{Paunescu}, and W.~J. {Stark},
  ``Robust chemical preservation of digital information on {DNA} in silica with
  error-correcting codes,'' \emph{Angewandte Chemie International Edition},
  vol.~54, no.~8, pp. 2552--2555, 2015.

\bibitem{HecShoRamTse17}
R.~{Heckel}, I.~{Shomorony}, K.~{Ramchandran}, and D.~N.~C. {Tse},
  ``Fundamental limits of {DNA} storage systems,'' in \emph{Proceedings of the
  2017 {IEEE} International Symposium on Information Theory ({ISIT}), Aachen,
  Germany}, Jun. 2017, pp. 3130--3134.

\bibitem{LenSieWacYaa19}
A.~{Lenz}, P.~H. {Siegel}, A.~{Wachter-Zeh}, and E.~{Yaakobi}, ``An upper bound
  on the capacity of the {DNA} storage channel,'' in \emph{Proceedings of the
  2019 {IEEE} Information Theory Workshop ({ITW}), Visby, Sweden}, Aug. 2019.

\bibitem{LevYaa19}
M.~{Levy} and E.~{Yaakobi}, ``Mutually uncorrelated codes for {{DNA}}
  storage,'' \emph{IEEE Trans.~on Inform.~Theory}, vol.~65, no.~6, pp.
  3671--3691, Jun. 2019.

\bibitem{LomQuiSim15}
N.~J. {Loman}, J.~{Quick}, and J.~T. {Simpson}, ``A complete bacterial genome
  assembled de novo using only nanopore sequencing data,'' \emph{Nature
  Methods}, vol.~12, no.~8, pp. 733--735, Aug. 2015.

\bibitem{MotRamTseMa13}
A.~{Motahari}, K.~{Ramchandran}, D.~{Tse}, and N.~{Ma}, ``Optimal {DNA} shotgun
  sequencing: Noisy reads are as good as noiseless reads,'' in
  \emph{Proceedings of the 2013 {IEEE} International Symposium on Information
  Theory ({ISIT}), Istanbul, Turkey}, Jul. 2013, pp. 1640--1644.

\bibitem{MotBreTse13}
A.~S. {Motahari}, G.~{Bresler}, and D.~N.~C. {Tse}, ``Information theory of
  {DNA} shotgun sequencing,'' \emph{IEEE Trans.~on Inform.~Theory}, vol.~59,
  no.~10, pp. 6273--6289, Oct. 2013.

\bibitem{NasShoVah22}
S.~{Nassirpour}, I.~{Shomorony}, and A.~{Vahid}, ``Reassembly codes for the
  chop-and-shuffle channel,'' \emph{arXiv preprint arXiv:2201.03590}, 2022.

\bibitem{OrgAngCheetal18}
L.~{Organick}, S.~D. {Ang}, Y.-J. {Chen} \emph{et~al.}, ``Random access in
  large-scale {DNA} data storage,'' \emph{Nature Biotechnology}, vol.~36,
  no.~3, pp. 242--248, Mar. 2018.

\bibitem{RavVahSho21}
A.~N. {Ravi}, A.~{Vahid}, and I.~{Shomorony}, ``Capacity of the torn paper
  channel with lost pieces,'' in \emph{Proceedings of the 2021 {IEEE}
  International Symposium on Information Theory ({ISIT}), Melbourne, Victoria,
  Australia}, Jul. 2021, pp. 1937--1942.

\bibitem{RavVahSho22}
------, ``Coded shotgun sequencing,'' \emph{{IEEE} Journal on Selected Areas in
  Information Theory}, vol.~3, no.~1, pp. 147--159, Mar. 2022.

\bibitem{Sal10}
S.~L. {Salzberg}, ``Mind the gaps,'' \emph{Nature Methods}, vol.~7, no.~2, pp.
  105--106, Feb. 2010.

\bibitem{ShoCouTse15}
I.~{Shomorony}, T.~{Courtade}, and D.~{Tse}, ``Do read errors matter for genome
  assembly?'' in \emph{Proceedings of the 2015 {IEEE} International Symposium
  on Information Theory ({ISIT}), Hong Kong, China}, Jun. 2015, pp. 919--923.

\bibitem{ShoHec19}
I.~{Shomorony} and R.~{Heckel}, ``Capacity results for the noisy shuffling
  channel,'' in \emph{Proceedings of the 2019 {IEEE} International Symposium on
  Information Theory ({ISIT}), Paris, France}, Jul. 2019, pp. 762--766.

\bibitem{ShoKamGovXiaCouTse16}
I.~{Shomorony}, G.~M. {Kamath}, F.~{Xia}, T.~A. {Courtade}, and D.~N. {Tse},
  ``Partial {DNA} assembly: A rate-distortion perspective,'' in
  \emph{Proceedings of the 2016 {IEEE} International Symposium on Information
  Theory ({ISIT}), Barcelona, Spain}, Jul. 2016, pp. 1799--1803.

\bibitem{ShoVah21}
I.~{Shomorony} and A.~{Vahid}, ``Torn-paper coding,'' \emph{IEEE Trans.~on
  Inform.~Theory}, vol.~67, no.~12, pp. 7904--7913, Dec. 2021.

\bibitem{Spe77}
J.~{Spencer}, ``Asymptotic lower bounds for {Ramsey} functions,''
  \emph{Discrete Mathematics}, vol.~20, pp. 69--76, 1977.

\bibitem{VarTen65}
R.~R. {Varshamov} and G.~M. {Tenengolts}, ``Code correcting single asymmetric
  errors (in {Russian}),'' \emph{Automatika i Telemkhanika}, vol.~26, no.~2,
  pp. 288--292, 1965.

\bibitem{WeiMer22}
N.~{Weinberger} and N.~{Merhav}, ``The {DNA} storage channel: Capacity and
  error probability bounds,'' \emph{IEEE Trans.~on Inform.~Theory}, vol.~68,
  no.~9, pp. 5657--5700, Sep. 2022.

\bibitem{WonWonFoo03}
P.~C. {Wong}, K.-k. Wong, and H.~{Foote}, ``Organic data memory using the {DNA}
  approach,'' \emph{Communications of the ACM}, vol.~46, no.~1, pp. 95--98,
  Jan. 2003.

\bibitem{YehBarMarYaa23}
Y.~{Yehezkeally}, D.~{Bar-Lev}, S.~{Marcovich}, and E.~{Yaakobi}, ``Generalized
  unique reconstruction from substrings,'' \emph{IEEE Trans.~on
  Inform.~Theory}, 2023.

\bibitem{YehPol21}
Y.~{Yehezkeally} and N.~{Polyanskii}, ``On codes for the noisy substring
  channel,'' in \emph{Proceedings of the 2021 {IEEE} International Symposium on
  Information Theory ({ISIT}), Melbourne, Victoria, Australia}, Jul. 2021, pp.
  1700--1705.

\end{thebibliography}
\end{document}